\documentclass{aastex701}

\usepackage{amsmath}

\begin{document}

\title{Anomaly Hunter for Alerts ({\sc AHA}): Anomaly Detection in the ZTF Transient Alert Stream}

\author[0009-0000-8422-2222]{Leyla Iskandarli}
\affiliation{Breakthrough Listen, Department of Physics, University of Oxford, Oxford OX1 3RH, UK}
\email[show]{iskandarlileyla@gmail.com}  

\author{Chris J. Lintott} 
\affiliation{Department of Physics, University of Oxford, Oxford OX1 3RH, UK}
\email{chris.lintott@physics.ox.ac.uk}

\author[0000-0003-4823-129X]{Steve Croft}
\affiliation{Breakthrough Listen, Department of Physics, University of Oxford, Oxford OX1 3RH, UK}
\affiliation{SETI Institute, Mountain View, CA 94043, USA}
\affiliation{Berkeley SETI Research Center, University of California, Berkeley, Berkeley CA 94720, USA}
\email{scroft@berkeley.edu}

\author[0000-0002-0504-4323]{Heloise Stevance}
\affiliation{Department of Physics, University of Oxford, Oxford OX1 3RH, UK}
\email{hfstevance@gmail.com}

\author{Joshua Weston}
\affiliation{Astrophysics Research Centre, School of Mathematics and Physics, Queen’s University Belfast, Belfast BT7 1NN, UK}
\email{jweston04@qub.ac.uk}

\begin{abstract}
Modern time-domain surveys produce alert streams at a scale that makes exhaustive manual inspection infeasible, requiring automated methods to identify unusual transients for follow-up. In this work, we present an unsupervised anomaly detection pipeline applied to the ZTF alert stream using the Lasair broker. We define normal objects as SN~Ia, SN~II, and SN~Ib/c. Anomalous objects include (i) more exotic transients (AGN, TDEs, SLSNe, CVs, and nuclear transients) and (ii) supernova-labeled objects, either spectroscopically or by Lasair, with anomalous properties, such as incorrect or absent host associations, or non-supernova-like light curves. Our pipeline consists of three independently trained simple autoencoders operating on distinct alert stream data products: object features, triplet image cutouts, and light curves. Each model is trained on predominantly normal transients, and performance is assessed using the recall of exotic objects and the purity of all anomalous objects across both a spectroscopically classified held-out test set and the live alert stream. In the test set, performance is evaluated at a fixed rank corresponding to the top ten scoring candidates, while in the alert stream it is evaluated using an anomaly threshold defined from test set behavior. Across both settings, the algorithms consistently recover exotic transients and anomalous supernovae among their top-ranked candidates. Over 25 days of live alert stream application, we identify 87 unusual supernova candidates for follow-up. The overlap between anomalies flagged by different autoencoders in the test set is non-existent, and in the alert stream is small, with maximum overlap between any two algorithms being 11 objects. The framework is data-efficient, requiring only a few thousand training examples, making it well suited for early and ongoing application to the Rubin Observatory alert stream.
\end{abstract}

\keywords{
\uat{Light curves}{918} --- 
\uat{Neural networks}{1938} --- 
\uat{Outlier detection}{1934} --- 
\uat{Supernovae}{1668} --- 
\uat{Time series analysis}{1916} --- 
\uat{Transient detection}{1957}
}


\section{Introduction}

Wide-field optical time-domain surveys now operate at alert rates that exceed the capacity of human vetting and traditional rule based filtering. The Zwicky Transient Facility \citep[ZTF;][]{Bellm2019} alone generates on the order of $10^6$  alerts per night \citep{Patterson2019}, a number that will soon increase by an order of magnitude with the Vera C.\ Rubin Observatory \citep{Bellm2020, Hambleton2023}. These alert streams contain a mixture of scientifically interesting transients. We adopt a broad definition of transients as sources whose observables change on timescales from minutes to years, including supernovae (SN), tidal disruption events (TDEs), kilonovae, cataclysmic variables (CVs), and active galactic nuclei (AGN). 

The discovery and follow-up of rare transients has repeatedly reshaped time-domain astrophysics by revealing new progenitor channels, explosion physics, and the diversity of luminous outbursts \citep{Margutti2019, Gagliano2022, Jacobson-Galan2022, Kuncarayakti2023, Pierel2023}. TDEs, for example, provide direct probes of black hole demographics and accretion physics \citep{Komossa2015}, while kilonovae from compact object mergers connect transient astronomy to heavy element nucleosynthesis and multi-messenger astrophysics \citep{Metzger2020}. Ongoing surveys also continue to uncover nuclear outbursts whose physical origin is uncertain, including long duration ambiguous nuclear transients and the most luminous extreme nuclear transients \citep[e.g.][]{Wiseman2025, Hinkle2025}. The statistical power and early time coverage enabled by the Vera Rubin Observatory Legacy Survey of Space and Time (LSST) can transform each of these areas by supplying both large, homogeneous samples and the rarest, most informative events \citep{Abell2009, Bellm2020}. Efficient triage therefore requires automated methods that can operate in real time and adapt to the scale of modern surveys. Recent work has demonstrated the effectiveness of supervised, feature based systems for operational alert filtering \citep{Stevance2025}.

Machine learning anomaly detection provides a framework for identifying rare, unusual, or previously unseen phenomena within large alert streams \citep{Chandola2009, Li2022}. Broadly, anomaly detection methods may be supervised, relying on labeled examples of known classes, or unsupervised, learning the structure of the data without explicit class labels and flagging deviations from this learned distribution. Most prior work in transient anomaly detection has focused on unsupervised or semi-supervised approaches applied to light curve data. Isolation Forest based methods have been applied to real and simulated photometric light curves to identify anomalous transients, including rare supernova subtypes, and microlensing 
events \citep{Pruzhinskaya2019, Webb2020, Ishida2021}. Autoencoder based models have been used to learn latent representations of simulated transient light curves, often in combination with Isolation Forests, demonstrating strong performance in flagging rare transient classes, such as superluminous supernovae (SLSNe), pair-instability supernovae (PISNe) and intermediate-luminosity optical transients  \citep{Villar2021}. Semi-supervised approaches have also been explored, using neural-network or parametric models trained on known transient classes to flag deviations as anomalies in real time alert streams \citep{Muthukrishna2022}. These methods primarily identify kilonovae, TDEs, intermediate-luminosity transients, calcium-rich gap transients, and PISNe. More recently, anomaly detection has been extended to feature based representations derived from alert streams, using combinations of autoencoders, deep support vector data description models, and hierarchical frameworks to identify anomalous sources in real ZTF alerts \citep{Perez-Carrasco2023}, as well as hybrid neural-network architectures operating on photometric statistical features, host galaxy associations and image cutouts to isolate rare transient populations \citep{Aleo2024, Sheng2024}. Supervised approaches have also been proposed, using latent embeddings from deep transient classifiers combined with class-conditioned Isolation Forests \citep{Gupta2025}. 

There is also an opportunity to search for technosignatures, observable signals indicative of extraterrestrial technology, within optical alert streams using anomaly detection approaches. Recent work has demonstrated that real time alert brokers can already be used to identify candidate technosignatures through targeted searches for unusual spatial, temporal, and spatiotemporal behavior \citep{Gallay2025}. These efforts have shown that broker-level filtering (implemented as rule based Structured Query Language queries within alert brokers, combining catalog cross-matching and object classification with explicit cuts on data quantity, brightness, timing, sky position, and simple variability indicators) can efficiently reduce millions of nightly alerts to fewer than five high-priority candidates suitable for follow-up. In this context, technosignatures are expected to manifest as unusual variability whose physical origin is not readily explained by known astrophysical processes, placing them within the broader category of anomalies. While fully automated machine learning pipelines for technosignature discovery in optical alert streams have not yet been realized, anomaly detection methods provide a natural extension of these broker-level approaches by enabling the identification of a wider and less prescriptive class of anomalous behavior beyond selection cuts.

Alert packets distributed by brokers such as Lasair \citep{Smith2019} contain multiple data products, each encoding different aspects of a transient. These include summary features derived from photometry, small image cutouts from difference imaging, and light curves that evolve as additional detections are reported. Each representation has distinct advantages and limitations. Feature vectors are compact and efficient, but may obscure morphological information and are sensitive to missing values. Image cutouts preserve spatial structure and host-galaxy context, but are more expensive to process and may be noisy at low signal-to-noise. Light curves encode temporal evolution, but are often sparse at early times. 

Rather than selecting a single representation, this work investigates anomaly detection independently across all three modalities to assess their relative sensitivity and the types of anomalies each reveals. Each model is trained on historical alert data and then applied both to held-out test sets and to a live alert stream. The resulting anomalies are examined in detail to identify the types of objects preferentially flagged by each modality and to assess the extent to which different representations recover overlapping or distinct subsets of unusual transients.  This paper is organized as follows. Section~\ref{sec:data} describes the data products used for algorithm training and testing. Section~\ref{sec:algorithm} defines anomalies and algorithm performance metrics, outlines the autoencoder architectures used for each data modality, and reports the algorithm performance. Section~\ref{sec:anomalies} applies the trained algorithms to the ZTF live alert stream and describes the interesting anomalies flagged. Section~\ref{sec:conclusion} discusses the results of this work and summarizes them.

The code is publicly available in Github,\footnote{\url{https://github.com/leylaiskandarli/AHA-Anomaly-Hunter-for-Alerts}} and the version of this code used in this work is available on Zenodo.\footnote{doi:\doi{10.5281/zenodo.18430441}}

\section{Data and Definitions \label{sec:data}}

ZTF \citep{Bellm2019} operates the Palomar 48-inch Schmidt telescope equipped with a 47\,deg$^{2}$ field-of-view camera and an 8\,s readout time, surveying the northern sky in the ZTF-$g$, ZTF-$r$, and ZTF-$i$\footnote{We exclude ZTF-$i$ observations from this analysis due to their limited cadence (approximately 10\% of ZTF observations).} passbands. In Phase~II operations, ZTF allocates 50\% of camera time and 50\% of the Spectral Energy Distribution Machine spectrograph time to a public survey of the entire northern sky on a two-night cadence. Image reduction and source detection are performed in near real time by the Infrared Processing and Analysis Center, with transient alerts generated from raw images within $\sim$4\,minutes \citep{Patterson2019}. These alerts are distributed publicly via the ZTF alert stream and made accessible through community alert brokers, including Lasair \citep{Williams2024}, ANTARES \citep{Matheson2021}, ALeRCE \citep{Forster2021}, Fink \citep{Moller2021}, AMPEL \citep{Nordin2019}, and Pitt-Google. These brokers are also designated to operate on the Rubin alert stream,  and their ZTF deployments serve as a precursor for Rubin. For this work, we use the Lasair alert broker.

We construct a labeled data set to train, validate, and test our anomaly detection pipeline using objects drawn from the full Lasair database of ZTF transients. We restrict this sample to sources with available spectroscopic classifications, obtained by cross-matching Lasair objects with the Transient Name Server (TNS) and supplemented by contextual classifications provided by the {\sc sherlock}\footnote{\url{https://zenodo.org/records/8289325}} \citep{Smith2020} framework. {\sc sherlock} assigns contextual classifications by cross-matching transients against a library of historical and ongoing sky-survey catalogs, enabling associations with nearby galaxies, known variable stars, AGN, or bright stars.

The resulting data set is intentionally dominated by normal supernovae (SN~Ia, SN~II, and SN~Ib/c), while objects not classified as these standard supernova types form a minority population. Throughout this work, we refer to this minority class as exotic objects. This includes sources labeled as AGN, TDEs, SLSNe, and nuclear transients by TNS and {\sc sherlock}. This imbalance is deliberate and defines which sources are considered anomalous in this work. The final data set contains 5294 unique objects: 3923 SN~Ia, 929 SN~II, 262 SN~Ib/c, and 186 exotic objects (see Table~\ref{tab:dataset}).


\begin{deluxetable*}{lcccc}
\tablecaption{Algorithm data set composition (5294 distinct objects) queried from Lasair. All have been spectroscopically classified.}
\label{tab:dataset}
\tabletypesize{\scriptsize}
\setlength{\tabcolsep}{6pt}
\tablehead{
\colhead{} &
\colhead{SN~Ia (3923)} &
\colhead{SN~II (929)} &
\colhead{SN~Ib/c (262)} &
\colhead{Exotic (186)}
}
\startdata
Subtypes (count) &
\begin{tabular}{@{}l@{}}Ia (3702)\\ Ia-91T-like (123)\\ Ia-91bg-like (43)\\ Ia-pec (18)\\ Iax[02cx-like] (17)\\ Ia-SC (11)\\ Ia-CSM (9)\end{tabular} &
\begin{tabular}{@{}l@{}}II (679)\\ IIn (138)\\ IIb (61)\\ IIP (45)\end{tabular} &
\begin{tabular}{@{}l@{}}Ic (104)\\ Ib (73)\\ Ic-BL (38)\\ Ib/c (16)\\ Ibn/Icn (9)\end{tabular} &
\begin{tabular}{@{}l@{}}SLSN-I (65)\\ SLSN-II (21)\\ AGN (55)\\ TDE (36)\\ TDE-H-He (4)\\ TDE-He (3)\\ TDE-featureless (2)\end{tabular} \\
\enddata
\end{deluxetable*}

From this data set, we derive three complementary data products. First, we use object features provided by Lasair, including statistical descriptors of the light curves and inferred host-galaxy properties obtained via {\sc sherlock}. Second, we retrieve alert based triplet image cutouts corresponding to each detection. Finally, we obtain full difference-flux light curves for each object directly from the broker. These data products are described in the following sections.

\subsection{Object features \label{sec:data_of}}

We gather object features that capture (i) the transient brightness, (ii) its recent photometric evolution, and (iii) its association with a predicted host galaxy. All features are summarized in Table~\ref{tab:lasair_ae_features}.

The use of statistical light curve summaries and host associations for anomaly detection has precedent in the literature. For example, \citet{Pruzhinskaya2019} and \citet{Ishida2021} applied algorithms to light curve features to identify rare or misclassified transients, while \citet{Aleo2024} used a similar class of photometric summary statistics and host context to characterize transient behavior and identify anomalous events within alert streams. Our approach is similar in spirit, but differs in that we restrict the input space to broker-native Lasair features, allowing us to assess the extent to which anomalies can be identified using only information available in real time, without costly or survey-specific post-processing.
 
Peak and typical magnitudes (e.g. {\it gmag}, {\it rmag}, {\it maggmin}, {\it magrmin}) provide an initial estimate of luminosity. The most recent change in magnitude divided by the time difference ({\it dmdt\_g}, {\it dmdt\_r}), together with exponential moving average magnitudes measured over 2, 8, and 28 days (e.g. {\it mag\_g02}, {\it mag\_g08}, {\it mag\_g28}; similarly in {\it r}), compress the light curve morphology into a small set of summary statistics sensitive to rapid rises, declines, or long-lived plateaus.

Additionally, host association metrics (e.g. {\it separationArcsec}, {\it major\_axis\_arcsec}, and {\it z}) are informative because some exotic transients occur at systematically different locations relative to host light: compact-object mergers can occur at large projected offsets due to natal kicks and long delay times, placing them well outside the bright stellar body of the host \citep[e.g.][]{Troja2019, Bom2025}. SLSNe also show host-offset distributions that differ from normal SNe, with an enhanced tail to larger host-normalized offsets in some samples \citep{Hsu2024}. In contrast, TDEs are associated with massive black holes and therefore expected to be coincident with galactic nuclei \citep[e.g.][]{Holoien2016}.

\begin{deluxetable}{ll}
\tablecaption{The 22 Lasair input features used for the object features autoencoder (See \href{https://lasair-ztf.lsst.ac.uk/schema/}{Lasair schema browser}).}
\label{tab:lasair_ae_features}
\tabletypesize{\scriptsize}
\setlength{\tabcolsep}{6pt}
\tablehead{
\colhead{Feature} & \colhead{Description}
}
\startdata
magrmin & Minimum $r$ magnitude of light curve (brightest). \\
magrmax & Maximum $r$ magnitude of light curve (faintest). \\
maggmean & Mean $g$ magnitude of light curve. \\
maggmin & Minimum $g$ magnitude of light curve (brightest). \\
maggmax & Maximum $g$ magnitude of light curve (faintest). \\
gmag & Latest $g$ magnitude. \\
rmag & Latest $r$ magnitude. \\
g\_minus\_r & Latest $(g-r)$ colour. \\
dmdt\_g & Most recent increase in $g$ magnitude divided by time difference (brightening $=$ positive); units: mag day$^{-1}$. \\
dmdt\_r & Most recent increase in $r$ magnitude divided by time difference (brightening $=$ positive); units: mag day$^{-1}$. \\
mag\_g02 & Latest exponential moving average of difference magnitude in $g$ band, with 2-day timescale. \\
mag\_g08 & Latest exponential moving average of difference magnitude in $g$ band, with 8-day timescale. \\
mag\_g28 & Latest exponential moving average of difference magnitude in $g$ band, with 28-day timescale. \\
mag\_r02 & Latest exponential moving average of difference magnitude in $r$ band, with 2-day timescale. \\
mag\_r08 & Latest exponential moving average of difference magnitude in $r$ band, with 8-day timescale. \\
mag\_r28 & Latest exponential moving average of difference magnitude in $r$ band, with 28-day timescale. \\
separationArcsec & Transient's angular separation (arcsec) from the top-ranked catalogue host match. \\
major\_axis\_arcsec & Size of associated galaxy in arcsec. \\
z & Spectroscopic redshift of host. \\
distpsnr1 & Distance to closest source from the PS1 catalogue (if one exists within 30 arcsec). \\
sgmag1 & $g$ band point-spread-function magnitude of closest source from the PS1 catalogue (if one exists within 30\,arcsec). \\
srmag1 & $r$ band point-spread-function magnitude of closest source from the PS1 catalogue (if one exists within 30\,arcsec). \\
\enddata
\end{deluxetable}

After selecting the input features, we apply conservative quality cuts to ensure that the autoencoder learns astrophysical structure rather than artifacts arising from non-detections or poorly constrained measurements. Objects with both $g$ and $r$ band magnitudes greater than 30\,mag are removed, as these values indicate non-detections or unreliable photometry in the stored summary statistics. We further exclude objects with large photometric-rate uncertainties by requiring the uncertainties in $dmdt\_g$ and $dmdt\_r$ to be less than 1. These cuts result in 3076 SN~Ia, 892 SN~II, 259 SN~Ib/c, and 172 exotic objects. For a subset of objects, Lasair does not provide an inferred host redshift ({\it z}) or galaxy size ({\it major\_axis\_arcsec}); in these cases we set the missing values to a null value (e.g. -999) so that “unknown host information” is represented consistently. Autoencoder based methods are commonly used in settings with missing or partially observed tabular data, and can learn representations that remain useful under missing information \citep[e.g.][]{Pereira2020}.

We require many normal objects while retaining a limited number of exotic objects for evaluation; however, we also want the model to see non-Ia supernovae during training. We therefore upsampled the minority normal classes (SN~II and SN~Ib/c) using the Synthetic Minority Over-sampling Technique \citep[SMOTE;][]{Chawla2002} until they matched the size of the majority normal class (SN~Ia). SMOTE synthesizes new samples by interpolating along line segments in feature space between each minority example and its $k$ nearest neighbors. We adopt $k=8$, equal to the smallest subtype count in our data set (SN~Ibn/Icn). The oversampling technique is similar to the approach taken by \citet{Aleo2024}. We ended up with a total of 9400 objects which included 3076 SN~Ia, 3076 SN~II, 3076 SN~Ib/c and 172 exotic objects. 

\subsection{Triplet image cutouts \label{sec:data_tc}}

Each alert is accompanied by a triplet of small image stamps centered on the candidate position: (i) the science image from the most recent visit, (ii) the template image constructed from prior observations, and (iii) the difference image produced by subtracting the template from the science image. These $64\times64$ pixel cutouts (with 1\,arcsec per pixel) are 32-bit Flexible Image Transport System (FITS) stamps that provide direct morphological information about the transient and the subtraction quality. The ZTF real time pipeline generates the difference image by performing image differencing (i.e. subtracting a reference image from the latest science image) using an optimized algorithm \citep{Zackay2016}. In the difference image, genuine transient sources appear as positive or negative detections, depending on whether the object brightened or faded relative to the reference.

For objects with multiple detections, we select a single representative triplet corresponding to the detection at which the source is brightest in the difference image. To quantify the significance of the transient signal in the selected difference cutouts, we compute a peak-pixel signal-to-noise ratio. The background noise, $\sigma$, is estimated using the median absolute deviation of pixel values after masking the transient signal, defined as the central three pixels of the stamp, to avoid bias from the source flux. The signal is then defined as the maximum absolute residual within this same central region, motivated by the fact that 83\% of peak residuals lie within three pixels of the stamp center. With this definition, the selected difference cutouts have a median peak-pixel signal-to-noise ratio of approximately 22, with a 16--84th percentile range of 9--56. Selecting the detection with the brightest difference image therefore minimizes the likelihood that the chosen cutout is dominated by low-significance subtraction residuals.

For each alert, the three image cutouts are stacked to form a three-channel input tensor of shape $3\times64\times64$ for each object. Prior to training, each cutout is standardized independently by subtracting the mean and dividing by the standard deviation computed over finite-valued pixels only. After standardization, non-finite pixel values are set to zero. This per-image normalization places the inputs in a relative intensity space, reducing sensitivity to absolute background levels and gain variations.

A practical limitation of using Lasair triplet cutouts is that image stamps are not retained indefinitely. Current Lasair guidance indicates that cutouts are stored for at least two years before being deleted (R. Williams, private communication). Consequently, we cannot recover triplets for the full spectroscopic data set. After applying these availability and selection criteria, our final triplet-image dataset contains 2786 objects: 2023 SN~Ia, 486 SN~II, 130 SN~Ib/c, and 147 exotic objects.

\subsection{Light curves \label{sec:data_light curve}}

We use light curves derived from the ZTF alert stream as served by the Lasair broker. Our goal is to isolate and compare the shape of transient brightness evolution for events that evolve on timescales typical of normal supernovae. 
For each object, we retrieve all available alert stream photometry from Lasair and retain only point-spread-function (PSF) difference magnitudes and their associated uncertainties in the $g$ and $r$ bands. These measurements are obtained from difference images. We do not use upper limits. 

Each light curve is aligned in time by defining $t=0$ as the brightest measured point in the raw photometry. The observations are then resampled onto a fixed time grid spanning 50 days before to 300 days after $t=0$, sampled every two days, resulting in 176 time steps. Resampling is performed independently in each band using weighted Gaussian kernel regression in magnitude space, where each grid value is computed as a weighted average of nearby observations. The kernel width is narrow near peak brightness and broader at later times, preserving sharp structure where the data are dense while providing smooth, conservative estimates in sparsely sampled regions.

To assess how well each resampled time step is constrained by the data, we compute a local support value from the kernel weights, representing the effective number of contributing observations at that time. Time steps supported by fewer than two effective measurements are masked and excluded from the representation. This requirement ensures that retained values are constrained by multiple observations rather than isolated measurements.

Objects with fewer than ten total detections are removed. Magnitudes in each band are scaled using the median, and median absolute deviation computed over all objects in the dataset. Masked time steps are set to zero after scaling so that missing data do not introduce a constant offset. After resampling and masking, each object is represented as a fixed-size tensor of 2$\times$176 with two channels corresponding to the $g$ and $r$ bands.

After applying all quality cuts, the final light curve dataset contains 2987 objects: 2015 SN~Ia, 694 SN~II, 122 SN~Ib/c, and 156 exotic transients.

\section{Algorithms \label{sec:algorithm}}

Autoencoders compress high-dimensional input data into a lower-dimensional latent space while preserving salient features of the original distribution \citep{Hinton2006}. For anomaly detection, the reconstruction loss produced by the autoencoder serves as a quantitative metric for identifying unusual objects. Additionally, the learned latent space representation enables exploration of the data manifold. 

We develop three independent autoencoder based anomaly detection algorithms, each operating on a single data modality and producing an anomaly score for every object. The network architectures are simple and distinct, adapted to the specific structure of each modality. Using a simple architecture reduces sensitivity to model-specific biases and limits overfitting to the training data, allowing differences in the resulting anomaly populations to be attributed primarily to the input data itself.

This section is structured as follows. We first define the anomalous objects of interest and the metrics used to evaluate algorithm performance. We then describe the algorithm constructed for each data modality, and evaluate each algorithm on a held-out test set drawn from the data set described in Section~\ref{sec:data}. These test sets preserve the overall class distribution of the data set, which is dominated by normal transients. Finally, we define an anomaly threshold based on the reconstruction loss distribution in the test set, and apply this fixed threshold when scoring objects in the live ZTF alert stream in the subsequent section.

\subsection{Definition of anomalous objects and evaluation metrics}

A clear definition of what constitutes an anomaly is essential for both the design of our pipeline and the evaluation of its performance. In this work, we consider two broad categories of anomalous objects.

The first category consists of exotic transients. We define exotic transients as all objects that are not spectroscopically classified as SN~Ia, SN~II, or SN~Ib/c. This category includes  AGN, SLSNe, TDEs, CVs, and nuclear transients. In addition, objects that have no spectroscopic classification, but are not classified by {\sc sherlock} as supernovae are also included in this exotic class.

The second category consists of objects that are labelled as supernovae (either spectroscopically or by {\sc sherlock}), but exhibit anomalous properties inconsistent with standard supernova behaviour. These include, for example, incorrect or ambiguous host-galaxy associations, large offsets from the host, the absence of a detectable host, unusually high brightness, evolution on multi-year timescales, or light curve morphologies that deviate from those expected for normal supernovae. 

Given these two anomaly classes, we require metrics that quantify both (i) how effectively exotic transients are recovered and (ii) how clean the set of flagged anomalies is when considering both exotic objects and anomalous supernovae. We therefore employ recall and purity, with each metric quantifying a different aspect of algorithm performance.

Recall is evaluated only for exotic transients and measures the fraction of exotic objects successfully flagged as anomalous by the algorithm. Purity measures the fraction of flagged objects that are genuinely interesting anomalies, including both exotic transients and anomalous supernovae.

Both metrics are evaluated at rank k, where k corresponds either to the top ten highest-scoring objects in the held-out spectroscopically labelled test set or to the anomaly threshold adopted for application to the alert stream.

The formal definitions are as follows:

\begin{equation}
\text{Exotic Object Recall}
= \frac{\mathrm{TP}_{\mathrm{Exo}}}{\mathrm{TP}_{\mathrm{Exo}} + \mathrm{FN}}
= \frac{\text{Exotic objects flagged}}{\text{All exotic objects}},
\end{equation}

\begin{equation}
\text{Purity}
= \frac{\mathrm{TP}}{\mathrm{TP} + \mathrm{FP}}
= \frac{\text{All anomalous objects flagged}}{\text{All objects flagged}}.
\end{equation}

For recall, true positives (TP$_{Exo}$) are exotic objects flagged as anomalous by the autoencoder, defined as having a reconstruction loss above the anomaly threshold, while false negatives (FN) are exotic objects not flagged as anomalous. For purity, true positives (TP) are either exotic objects or anomalous supernova-labelled objects flagged by the autoencoder, while false positives (FP) correspond to normal supernovae incorrectly flagged as anomalous.

\subsection{Autoencoder 1: object features \label{sec:algorithm_features}}

The dataset, described in Section~\ref{sec:data_of}, is partitioned into training, validation, and test subsets in an 80/10/10 ratio. The autoencoder architecture was implemented using the {\sc Keras} package \citep{Chollet2018}. A fully connected, symmetric encoder and decoder were used. The encoder contains three hidden layers with widths 88, 56, and 27, followed by a latent space of dimension 8. The decoder mirrors this structure to the output. The sizes 88, 56, 27, and 8 refer to internal layer dimensions, not the input size. The six latent variables summarize joint variation across all 22 inputs (See Table~\ref{tab:lasair_ae_features}) and allow nonlinear combinations. Each latent combines information from multiple inputs, and each input is reconstructed from all latent dimensions through the learned decoder weights. Rectified Linear Unit activations were applied at each hidden layer, and a final sigmoid activation was used in the output layer to produce normalized reconstructions. Batch normalization was applied after each activation layer to stabilize training. All features were scaled to the range $0\leq X^{\prime}_i\leq 1$, where $X^{\prime}_i$ is a single input feature vector, in order to match the sigmoid activation function applied in the output layer of the decoder. This choice of normalization was motivated by empirical findings in previous studies that demonstrate improved performance when using bounded activation functions on scaled input data \citep{Amorim2022}. The model was optimized with the Adam optimizer \citep{Kingma2014}. 

The Huber loss function was employed during training and is defined as:

\begin{equation}
L_\delta(x, \hat{x}) =
\begin{cases}
\frac{1}{2}(x - \hat{x})^2 & \text{for } |x - \hat{x}| \leq \delta, \\
\delta \left( |x - \hat{x}| - \frac{1}{2} \delta \right) & \text{otherwise}
\end{cases}
\end{equation}

where \(\delta\) sets the transition between quadratic and linear behavior. It is quadratic for small residuals and linear for large residuals, combining the advantages of mean squared error and mean absolute error. This loss is adopted to better handle extreme values, which are often poorly reconstructed under mean squared error, a known issue in autoencoders \citep{Neloy2024}. Using the Huber loss mitigates these effects and improves robustness when reconstructing extreme-valued inputs. Training was stopped at a validation set reconstruction loss of approximately $10^{-5}$.

\begin{figure}
\centering
\includegraphics[width=\linewidth]{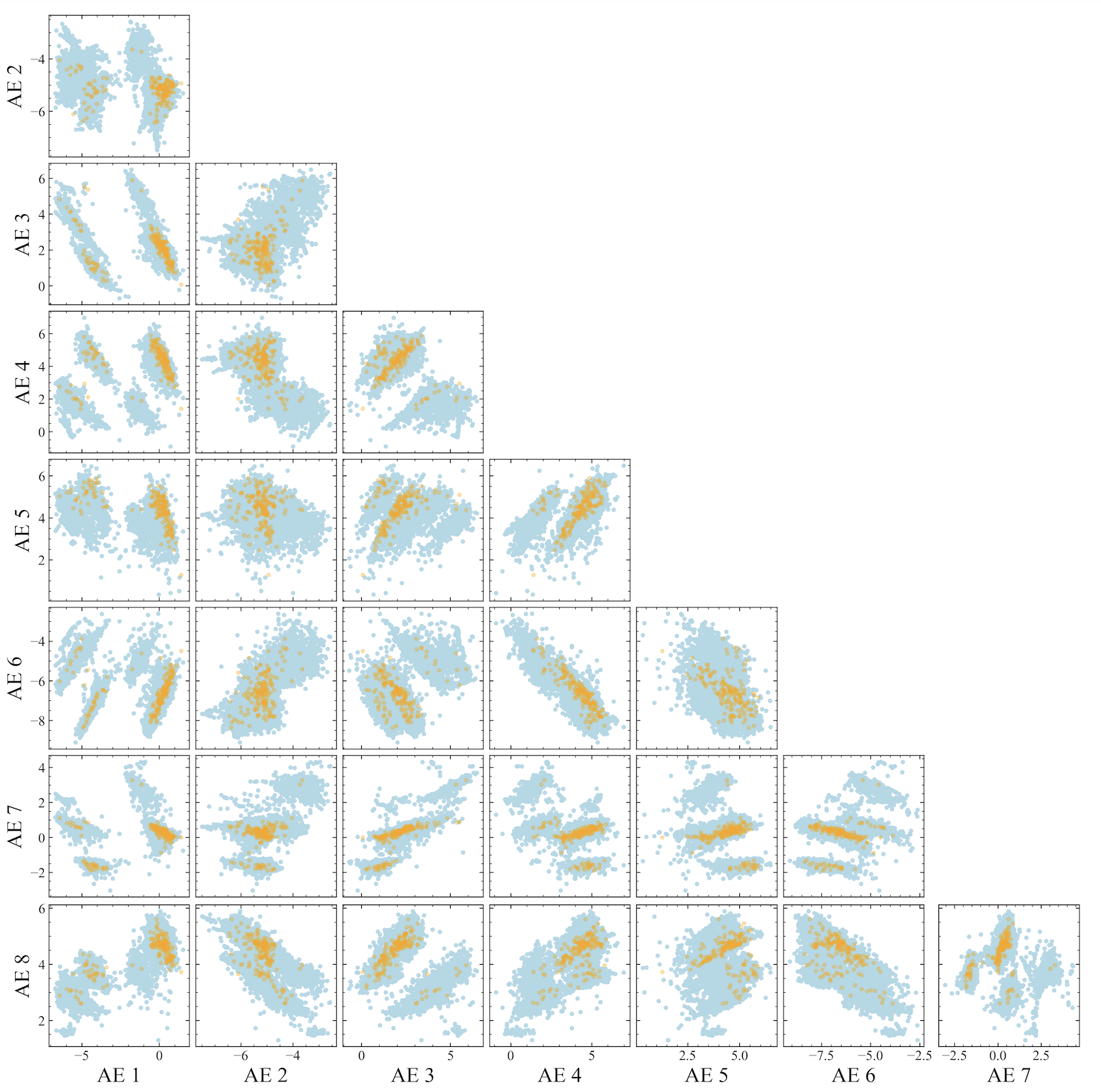}
\caption{Latent space projections of the autoencoder bottleneck dimensions. Axes are arbitrary latent dimensions. Normal objects are shown in blue and exotic objects are shown in orange. Exotic objects occupy overlapping regions of latent space and form localized substructure within the broader distribution rather than a clearly separated population.}
\label{fig:latentspace}
\end{figure}

In Figure~\ref{fig:latentspace}, the latent space representation of the full object features dataset is shown, with exotic objects marked in orange and normal objects marked in blue. Exotic objects form local concentrations within the latent space, but substantial overlap with the normal-object distribution is present. The absence of clear separation in two-dimensional latent-space projections may indicate that distinctions between normal and exotic objects are encoded in higher-dimensional relationships that are not captured in low-dimensional views, and only emerge when the full latent representation is considered jointly. This provides a further motivation for the use of an autoencoder, which is well suited to identifying anomalous structure in high-dimensional data \citep{Hinton2006}.

\begin{figure*}[t]
\centering
\includegraphics[width=\linewidth]{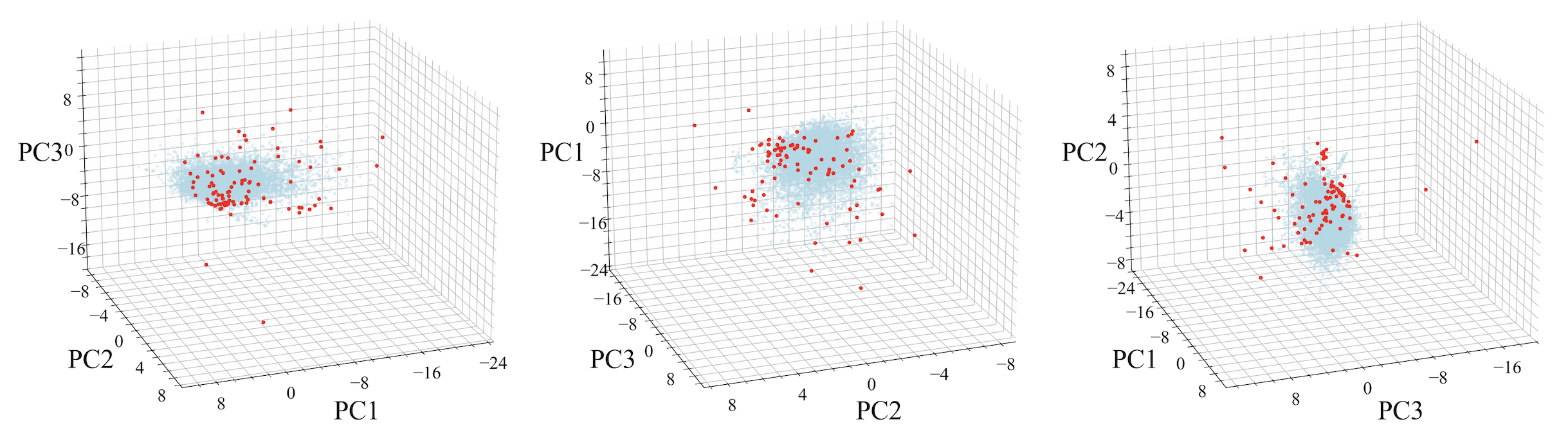}
\caption{Three-dimensional principal component projections of the full dataset. Blue points represent the full object population, while red points indicate the top 1\% of objects ranked by autoencoder reconstruction error. Each panel shows the same data from a different viewing angle to illustrate the spatial distribution of high-reconstruction-error objects relative to the PCA manifold.}
\label{fig:PCA_all}
\end{figure*}

We next consider the three-dimensional principal component analysis (PCA) projection of the full object-features dataset shown in Figure~\ref{fig:PCA_all}, with the top 1\% of objects ranked by autoencoder reconstruction error highlighted in red. PCA is used here as a simple, linear baseline to identify obvious outliers in a low-dimensional projection of the high-dimensional feature space and to assess whether these same objects are recovered by the autoencoder. Several features are evident. A small number of extreme PCA outliers is present, associated with incorrect host-galaxy associations. These cases are consistent with a known systematic in which {\sc sherlock} associates sources with nearby dwarf galaxies despite unrealistically large transient-host separations, implying an incorrect host assignment rather than a physical association (S. Smartt, private communication). Although the affected subset is small, identification is desirable because such failures can bias downstream feature construction and classification. In addition, high reconstruction-error objects are located within the high-density regions of the PCA manifold. This behavior is consistent with the autoencoder exploiting multivariate feature relationships that are not apparent in low-dimensional linear spaces.

Evaluation on the held-out test set is more informative, as these objects were not used during training. We therefore evaluate ten test set objects with the highest reconstruction loss. The test set contains 43 exotic objects, implying a maximum possible exotic object recall of 23\% at top ten. One AGN (AT~2019aawv) gets flagged, corresponding to an exotic object recall of 2.3\% at top ten. Purity is evaluated separately by inspecting the objects flagged as anomalous more carefully. Within the top ten, several objects classified as SN~Ia and SN~II exhibit feature values that are inconsistent or incorrect. These include missing host associations (SN~2022lxg), incorrect host associations (SN~2023jxe, SN~2022adcg), unclear host associations (SN~2025arz), and misclassification by {\sc sherlock} as either a variable star (SN~2025pht) or a nuclear transient (SN~2019xj). We treat these sources as true anomalies in the context of this study, despite their normal spectroscopic classifications.

The remaining three objects in the top ten have faint host associations (SN~2024kng, SN~2024egg) or correspond to an otherwise typical SN~II event (SN~2023cnd), and are considered false anomalies. Under this broader definition of anomalous objects, including both labeled exotics and sources with anomalous feature values, the purity at the top ten highest reconstruction loss is 70\%. In the test set, purity is the more informative metric, as it quantifies how many genuinely anomalous objects are present among the ten highest ranked candidates.

We select an anomaly threshold of 0.02, corresponding to the lowest reconstruction loss among true positives within the top ten. This threshold is adopted for application to the live alert stream (see Section~\ref{sec:feature_anoms}).

\subsection{Autoencoder 2: triplet image \label{sec:algorithm_tc}}

For the cutout images described in Section~\ref{sec:data_tc}, we adopt a convolutional autoencoder. This architecture is a direct analogue of the fully connected autoencoder used for the other modalities, but replaces dense layers with convolutional filters that are better suited to capturing the local spatial structure of image data \citep{Masci2011}

The autoencoder consists of a symmetric encoder–decoder architecture with three convolutional layers in each branch. Down sampling is performed using stride-2 convolutions, while up sampling is achieved with transposed convolutions. The encoder applies convolutional layers with kernel size 3, padding 1, and stride 2, mapping channel depth as $3\rightarrow32\rightarrow64\rightarrow128$ and reducing spatial dimensions as $64\rightarrow32\rightarrow16\rightarrow8$. The resulting latent representation has shape $128\times8\times8$. The decoder mirrors this structure and restores the spatial dimensions to $64\times64$. Rectified Linear Unit activations are used after each intermediate layer, and no output activation is applied, as the reconstruction target is defined in standardized pixel space. The model is trained using a mean squared error reconstruction loss.

\begin{figure}
    \centering
    \includegraphics[width=\linewidth]{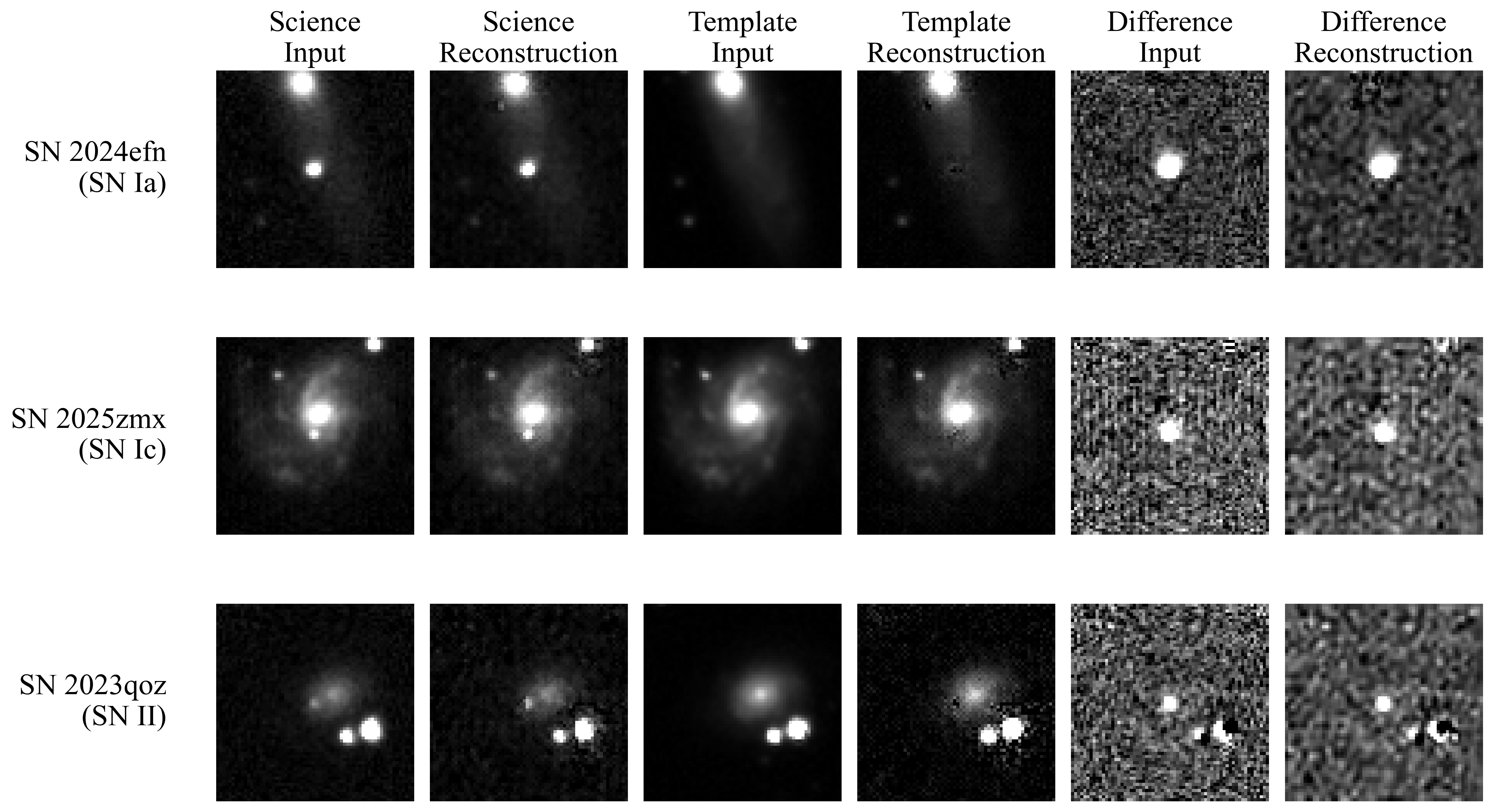}
    \caption{Input triplet images (science, template, difference) and corresponding autoencoder reconstructions for well-reconstructed objects, illustrating the model’s ability to reproduce typical alert cutout morphology.}
    \label{fig:TCs}
\end{figure}

The dataset is split into training, validation, and test subsets using an 80/10/10 split. After training, the autoencoder achieves a mean reconstruction loss of approximately $10^{-3}$ on the validation set. Examples of objects that are reconstructed well by the triplet autoencoder are shown in Figure~\ref{fig:TCs}. We then examine the objects ranked in the top 1\% by reconstruction error across the full dataset. These are most commonly sources with noisy or low signal-to-noise cutouts, particularly where the template image is itself noisy, which is often indicative of a faint or undetected host galaxy. This behavior is consistent with prior work, in which a convolutional neural network trained on both image cutouts and contextual information, including transient brightness and host-galaxy properties, preferentially identified SLSNe and TDEs from the ZTF Bright Transient Survey sample \citep{Sheng2024}.

We apply the autoencoder to the held-out test set and examine the ten highest-scoring objects. These consist of nine SN~Ia objects and one SLSN-I. The test set contains ten exotic objects in total, setting a maximum possible exotic object recall of 100\% at top ten highest-scoring objects. Since only one exotic object (SN~2025whv) is flagged, the exotic object recall for this algorithm on the test set is 10\%. The SN~Ia objects are mainly associated with very faint hosts, no detected host (SN~2025izg, SN~2023tqm), or, in one case, an incorrect host association (SN~2025cxg). Supernovae with faint or undetected hosts are not considered sufficiently anomalous for our purposes. As a result, only four of the ten flagged objects are considered truly anomalous, giving a purity of 40\%. Notably, the anomalies recovered by our triplet image autoencoder in the test set are distinct from those identified by our object features autoencoder.

For the test set, the lowest reconstruction loss among the true anomalies is 0.07; we adopt this value as the anomaly threshold for image based alerts in the live alert stream (see Section~\ref{sec:triplet_anoms}).

\subsection{Autoencoder 3: light curve \label{sec:algorithm_light curve}}

For the light curve modality, we use a autoencoder trained on the fixed-grid, two-band light curve representations described in Section~\ref{sec:data_light curve}. The dataset is split into training, validation, and test subsets using an 80/10/10 ratio. The autoencoder adopts a symmetric encoder-decoder architecture with three hidden layers of dimension 256, 128, and 64, followed by a latent space of dimension 8. Rectified Linear Unit activations are applied after each hidden layer, and no activation is applied at the output layer, as the reconstruction target is defined in scaled magnitude space.

\begin{figure}
    \centering
    \includegraphics[width=\linewidth]{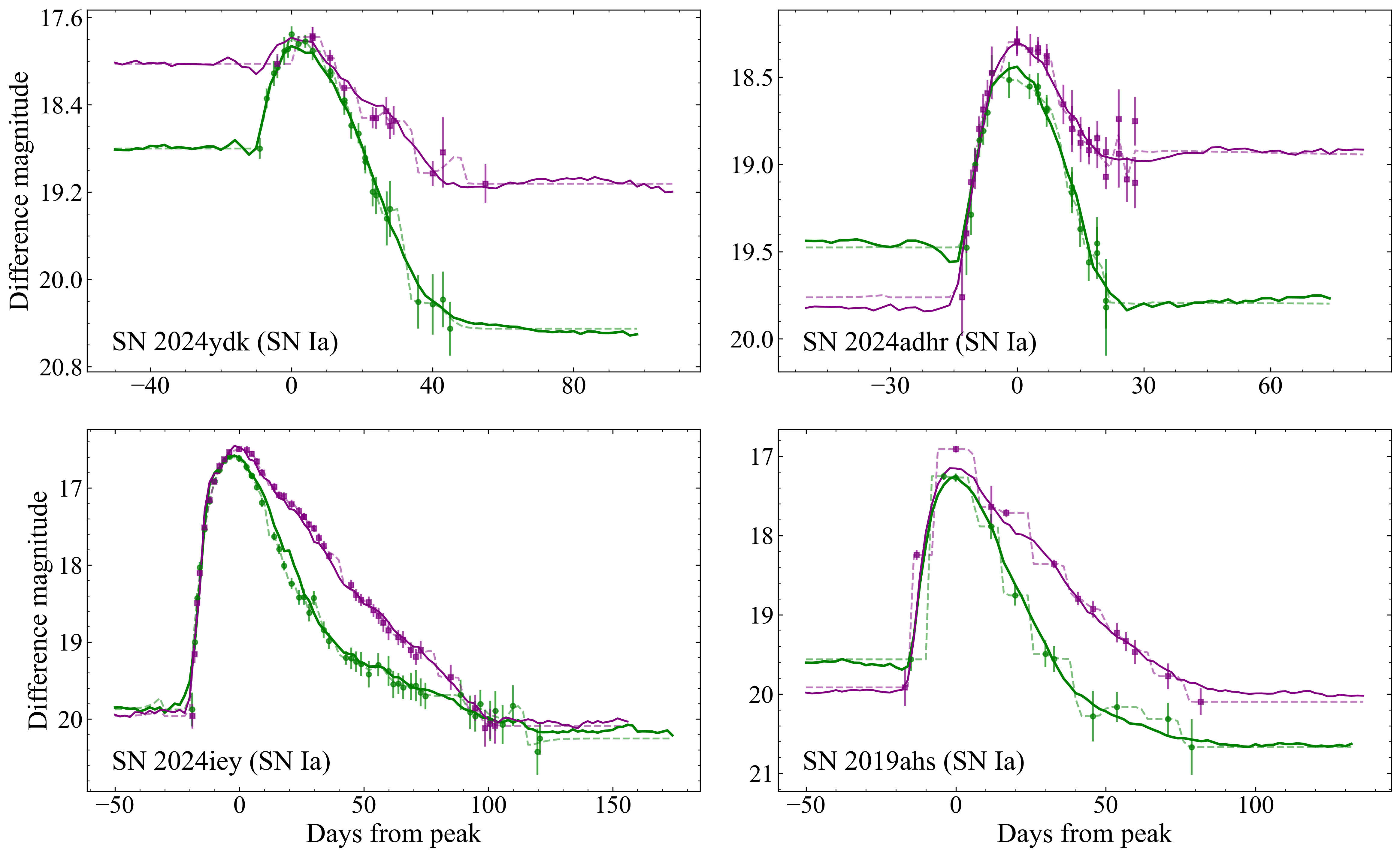}
    \caption{Example light curves from the training set used by the light curve autoencoder, shown after model training. All panels display normal SN~Ia objects. Green points and curves correspond to the $g$ band, and purple points and curves to the $r$ band. Discrete points with error bars show the raw ZTF difference-flux photometry retrieved from Lasair. The dashed, semi-transparent curves indicate the kernel-regressed light curves resampled onto a fixed temporal grid, which form the inputs to the autoencoder. Solid curves show the corresponding autoencoder reconstructions.}
    \label{fig:light curve_recon_examples}
\end{figure}

The model is trained using a weighted mean squared error reconstruction loss. Reconstruction errors are weighted by the effective support of each light curve time step, so that sparsely sampled phases contribute less to the loss than well-constrained regions. This weighting prevents light curves with limited coverage from being systematically assigned higher anomaly scores due to sparse sampling. Optimization is performed using the Adam optimizer \citep{Kingma2014}, and training proceeds for 500 epochs. After training, the mean reconstruction loss on the validation set converges to approximately 0.02. Visual inspection of reconstructed light curves confirms that the model accurately reproduces the typical rise-and-decline morphology of normal supernova light curves, as illustrated for representative SN~Ia training examples in Figure~\ref{fig:light curve_recon_examples}.

We next apply the trained autoencoder to the held-out test set and inspect the ten objects with the highest reconstruction loss. These objects are spectroscopically classified as one AGN, three SN~Ia, one SN~Ic, one SN~IIb, one SN~Ib, two SN~IIn, and one SN~Ia-CSM. The held-out test set contains 12 exotic objects, setting a maximum possible exotic object recall of 83\%. Since only one exotic object (AT~2019pcn) is flagged within the top ten, the exotic object recall at rank ten is 8.3\%.

Several of the supernova-labeled objects within the top ten exhibit anomalous properties. The most extreme case is SN~2020nps, classified as SN~IIn on TNS, but displaying a multi-year, still-rising light curve that is inconsistent with supernova evolution and more consistent with AGN variability. The transient position is coincident with the galaxy LEDA~4488494, which hosts an active galactic nucleus identified by the MaNGA survey \citep{Alban2023}. The angular separation between the transient and the galaxy nucleus is 0.21\,arcsec. The {\sc sherlock} classifier independently identifies this source as an AGN.

Other objects in the top ten include a transient with a double-peaked light curve (SN~2023aew), a rare SN~Ia-CSM event (SN~2022erq), a transient located far from its host galaxy (SN~2022actt), and an object with no detected host (SN~2025fmr). The remaining four objects are normal supernovae with broad light curve peaks or poorly constrained maxima. Overall, six of the ten flagged objects are considered anomalous, yielding a purity of 60\% for the held-out test set. None of the test set anomalies identified by the light curve autoencoder overlap with those identified by the object features or triplet image autoencoders in their respective test sets.

Based on the test set ranking, we adopt a reconstruction loss threshold of 0.2 for the light curve modality when applying the trained autoencoder to the live alert stream (see Section~\ref{sec:lightcurve_anoms}). Below this value, the highest-ranked objects are dominated by broadly peaked, but otherwise typical supernova light curves rather than the morphologies of interest.

\section{Alert Stream Anomalies \label{sec:anomalies}}

We have developed and tested autoencoder based anomaly detection algorithms for three complementary data modalities, showing that each identifies distinct anomalies. The object features autoencoder preferentially flags sources with incorrect or missing host-galaxy associations, and unusual multivariate feature relationships that are not apparent in low-dimensional projections. The triplet image autoencoder primarily identifies transients associated with faint, unresolved, or undetected host galaxies. The light curve autoencoder flags objects whose photometric evolution deviates from the characteristic rise-and-decline morphology of normal supernovae. In this section, we apply each trained algorithm to the live ZTF alert stream and analyze the flagged anomalies.

A public filter was constructed on the Lasair broker to access the ZTF alert stream for anomaly scoring. The filter was designed to return all objects classified as transients by the Lasair ingestion pipeline, with minimal astrophysical selection. No cuts were applied on brightness, color, host association, Galactic latitude, or external classifications, in order to maximize completeness and minimize bias against rare or poorly characterized events. Objects classified as variable stars were filtered out of the alert stream and are not included in the analysis.

The filter queries the “objects” table for transients whose most recent alert time falls within the chosen date range. Contextual information from {\sc sherlock} host associations and cross-matches to TNS are joined when available, but are not required for inclusion. Most alerts in the stream lack spectroscopic classifications, and object types are therefore primarily assigned by the {\sc sherlock} classifier.

The trained autoencoders, described in Section~\ref{sec:algorithm}, were applied to alerts returned by this filter between 2025-12-01 and 2025-12-25. 

\subsection{Object features autoencoder anomalies \label{sec:feature_anoms}}

In total, 206 objects are flagged by the object features autoencoder. Using contextual classifications provided by {\sc sherlock}, these comprise 95 supernovae, 86 AGN, ten nuclear transients, seven CV, four orphan transients with no associated host galaxy, and four objects with unclear host associations.

Alerts assigned high anomaly scores typically exhibit reconstruction losses two to three orders of magnitude above the anomaly threshold of 0.02, which was defined in Section~\ref{sec:algorithm_features}. Objects with the largest reconstruction errors are primarily characterized by poorly constrained photometric rate features, most notably {\it dmdt\_g} and {\it dmdt\_r}. These quantities measure the most recent change in $g$ and $r$ band PSF fit magnitude divided by the time difference. As a result, the prevalence of AGN, CVs, and nuclear transients among the anomalous population can be largely driven by extreme values of {\it dmdt\_g} and {\it dmdt\_r}, which manifest as sharp flux changes in their light curves.

We then perform a manual, object-by-object inspection of the candidates classified as supernovae, orphan transients, and sources with unclear host associations, and assign a qualitative priority rank based on scientific interest. These objects and their ranks are listed in Table~\ref{tab:features_ae_anoms1}. Ranks range from 1 to 3. Rank~3 objects are the highest priority: they lack spectroscopic classifications and exhibit unusual light curve behavior spanning multiple years. Rank~2 objects are potentially interesting, but have not yet evolved sufficiently for confident interpretation and also lack spectroscopic classification. Rank~1 objects are lower priority, either because they have already been spectroscopically classified, have been detected too recently to allow meaningful interpretation, or appear to be normal supernovae. This manual ranking is used to prioritize sources for follow-up and to inform future iterations of the anomaly detection pipeline.

\subsubsection{Most interesting objects: rank 3}

Objects assigned rank~3 exhibit light curves evolving over multi-year timescales and can be associated with either faint host galaxies or no detected host. We assign rank~3 to 16 objects. Most objects in this category are listed on TNS as possible supernovae and lack multi-wavelength detections.\footnote{Our search consisted of coordinate queries in SIMBAD (and NED), and cross-matches against {\em Gaia}, Pan-STARRS, {\em AllWISE}, 2MASS, {\em GALEX}, {\em ROSAT} (and {\em XMM-Newton}/{\em Swift}/{\em eROSITA} where available), as well as the radio survey NVSS.} Exceptions include AT~2024rsq, which has been recommended for spectroscopic follow-up by \citet{Xu2024b} on TNS as a possible nova; AT~2023qhf, which has a radio counterpart \citep{Vigotti1989} and has been reported as an AGN on TNS, specifically a blazar\footnote{\url{https://www.virtualtelescope.eu/2023/08/16/discovery-of-variability-of-the-optical-counterpart-of-the-nvss-j004354404634-radio-source-in-the-messier-31-field/}}. AT~2017cox is reported as a possible CV on TNS \citep{Prieto2017}. AT~2016dum has a cataclysmic binary reported at its coordinates \citep{Kawash2021}, and is most likely a CV. ALeRCE predicts AT~2022fvo to be a quasar, AT~2021adsf and AT~2024qs to be SLSNe, and AT~2021adir to be an AGN.

\subsubsection{Moderately interesting objects: rank 2}

Objects assigned rank 2 exhibit non-supernova-like light curves and evolve over shorter timescales. These objects can also have faint or ambiguous host associations. Light curves inconsistent with typical supernova evolution are more suggestive of CVs, variable stars, or AGN, although spectroscopic follow-up remains necessary in the absence of radio counterparts. Most objects in this category are reported on TNS as possible supernovae. Notable cases include AT~2023cqq, whose coordinates coincide with an OH/IR star and multiple infrared sources \citep{Cutri2003}, suggesting a crowded stellar environment; and AT~2018hgp, which aligns with a cataclysmic binary candidate \citep{Szkody2020} and exhibits a light curve consistent with a CV. AT~2025ojf is predicted to be a quasar by ALeRCE, and is centrally located within its host galaxy. The coordinates of AT~2019fex coincide with a known eclipsing binary \citep{Chen2020}, consistent with its light curve.

\subsubsection{Least interesting objects: rank 1}
Objects assigned rank 1 were not considered interesting enough for further investigation as they were either normal supernova-like objects or their light curves were too sparse to make any meaningful assumptions. These objects were primarily flagged as anomalous due to known systematics, including incorrect host-galaxy associations, the absence of a detected host galaxy, and nonphysically large {\it dmdt\_g} or {\it dmdt\_r} values driven by sparse sampling or large photometric uncertainties. In cases where the light curves are sparse, continued monitoring by all-sky surveys provides a mechanism to track their evolution should they later develop features indicative of more interesting behavior.

\begin{deluxetable}{lllll}
\tablecaption{Alert stream anomalies flagged by the object features autoencoder}
\label{tab:features_ae_anoms1}
\tabletypesize{\scriptsize}
\tablehead{
\colhead{ZTF ID} & \colhead{IAU Name} & \colhead{{\sc sherlock} Classification} & \colhead{Notes} & \colhead{Rank}
}
\startdata
\href{https://lasair-ztf.lsst.ac.uk/objects/ZTF21abashma/}{ZTF21abashma}	&	2021aibi	&	SN	&	800 day rise. Non-SN-like light curve	&	3	\\
\href{https://lasair-ztf.lsst.ac.uk/objects/ZTF20abnvwor/}{ZTF20abnvwor}	&	2025vmp	&	SN	&	800 day rise. Non-SN-like light curve	&	3	\\
\href{https://lasair-ztf.lsst.ac.uk/objects/ZTF19aburnjy/}{ZTF19aburnjy}	&	2022dxn	&	SN	&	800 day evolution. Non-SN-like light curve	&	3	\\
\href{https://lasair-ztf.lsst.ac.uk/objects/ZTF20abpehlo/}{ZTF20abpehlo}	&	2023qhf	&	SN	&	800 day evolution. Far from host. Non-SN-like light curve	&	3	\\
\href{https://lasair-ztf.lsst.ac.uk/objects/ZTF19aagmutb/}{ZTF19aagmutb}	&	2025ahll	&	SN	&	800 day evolution. Far from host. Non-SN-like light curve	&	3	\\
\href{https://lasair-ztf.lsst.ac.uk/objects/ZTF22aagawty/}{ZTF22aagawty\tablenotemark{$\dagger$}}	&	2022fvo	&	SN	&	800 day evolution. Faint host. Non-SN-like light curve	&	3	\\
\href{https://lasair-ztf.lsst.ac.uk/objects/ZTF20achtkxi/}{ZTF20achtkxi}	&	2020aezk	&	SN	&	800 day  rise. Non-SN-like light curve	&	3	\\
\href{https://lasair-ztf.lsst.ac.uk/objects/ZTF21acjcarx/}{ZTF21acjcarx\tablenotemark{$\star$}}	&	2021adsf	&	SN	&	800 day evolution. Non-SN-like light curve	&	3	\\
\href{https://lasair-ztf.lsst.ac.uk/objects/ZTF20acxqxyb/}{ZTF20acxqxyb}	&	2021eci	&	SN	&	700 day evolution. Non-SN-like light curve	&	3	\\
\href{https://lasair-ztf.lsst.ac.uk/objects/ZTF23abscfuu/}{ZTF23abscfuu}	&	2023aclf	&	SN	&	700 day evolution. Non-SN-like light curve	&	3	\\
\href{https://lasair-ztf.lsst.ac.uk/objects/ZTF23aagdltf/}{ZTF23aagdltf\tablenotemark{$\dagger$}}	&	2024qs	&	SN	&	700 day evolution. Non-SN-like light curve	&	3	\\
\href{https://lasair-ztf.lsst.ac.uk/objects/ZTF18aabezwv/}{ZTF18aabezwv}	&	2016dum	&	SN	&	700 day evolution. Faint host. Non-SN-like light curve	&	3	\\
\href{https://lasair-ztf.lsst.ac.uk/objects/ZTF19aaklfkk/}{ZTF19aaklfkk\tablenotemark{$\star$}}	&	2017cox	&	SN	&	600 day evolution. Far from host. Non-SN-like light curve	&	3	\\
\href{https://lasair-ztf.lsst.ac.uk/objects/ZTF21acldimi/}{ZTF21acldimi\tablenotemark{$\star$}}	&	2021adir	&	SN	&	600 day evolution. Unclear host. Non-SN-like light curve	&	3	\\
\href{https://lasair-ztf.lsst.ac.uk/objects/ZTF24aakccfa/}{ZTF24aakccfa}	&	2023aclf	&	SN	&	600 day evolution. Non-SN-like light curve	&	3	\\
\href{https://lasair-ztf.lsst.ac.uk/objects/ZTF24aatkhkt/}{ZTF24aatkhkt}	&	2024rsq	&	SN	&	500 day rise. Far from host. Non-SN-like light curve	&	3	\\
\href{https://lasair-ztf.lsst.ac.uk/objects/ZTF20aaaxdvw/}{ZTF20aaaxdvw}	&	2020hs	&	ORPHAN	&	Non-SN-like light curve. No host	&	2	\\
\href{https://lasair-ztf.lsst.ac.uk/objects/ZTF25aagpzug/}{ZTF25aagpzug}	&	2025cie	&	ORPHAN	&	Non-SN-like light curve. No host	&	2	\\
\href{https://lasair-ztf.lsst.ac.uk/objects/ZTF20aahbzhi/}{ZTF20aahbzhi}	&	2018krw	&	UNCLEAR	&	Non-SN-like light curve. Unclear host	&	2	\\
\href{https://lasair-ztf.lsst.ac.uk/objects/ZTF21aaanzzg/}{ZTF21aaanzzg\tablenotemark{$\dagger$}}	&	2021acf	&	UNCLEAR	&	Non-SN-like light curve. Unclear host	&	2	\\
\href{https://lasair-ztf.lsst.ac.uk/objects/ZTF18adohecj/}{ZTF18adohecj\tablenotemark{$\star$}}	&	2016hfm	&	SN	&	Non-SN-like light curve. Far from host 	&	2	\\
\href{https://lasair-ztf.lsst.ac.uk/objects/ZTF18abfxhsz/}{ZTF18abfxhsz}	&	2024fnn	&	SN	&	Non-SN-like light curve. Far from host	&	2	\\
\href{https://lasair-ztf.lsst.ac.uk/objects/ZTF25aacdief/}{ZTF25aacdief}	&	2023acmv	&	SN	&	Non-SN-like light curve. Far from host	&	2	\\
\href{https://lasair-ztf.lsst.ac.uk/objects/ZTF18acarunx/}{ZTF18acarunx\tablenotemark{$\star$}}	&	2018hgp	&	SN	&	Non-SN-like light curve. Faint host	&	2	\\
\href{https://lasair-ztf.lsst.ac.uk/objects/ZTF25aawdyvn/}{ZTF25aawdyvn}	&	2025ojf	&	SN	&	Non-SN-like light curve. Faint host 	&	2	\\
\href{https://lasair-ztf.lsst.ac.uk/objects/ZTF18aczrwbr/}{ZTF18aczrwbr\tablenotemark{$\star$}}	&	2023cqq	&	SN	&	Non-SN-like light curve. Faint host	&	2	\\
\href{https://lasair-ztf.lsst.ac.uk/objects/ZTF18adatjjz/}{ZTF18adatjjz}	&	2021uzg	&	SN	&	Non-SN-like light curve. Faint host	&	2	\\
\href{https://lasair-ztf.lsst.ac.uk/objects/ZTF18aabfxpw/}{ZTF18aabfxpw}	&	2018kfz	&	SN	&	Non-SN-like light curve. Faint host	&	2	\\
\href{https://lasair-ztf.lsst.ac.uk/objects/ZTF18acaujas/}{ZTF18acaujas}	&	2019aaey	&	SN	&	Non-SN-like light curve. Faint host	&	2	\\
\href{https://lasair-ztf.lsst.ac.uk/objects/ZTF18aaaatda/}{ZTF18aaaatda}	&	2020aadz	&	SN	&	Non-SN-like light curve. Faint host	&	2	\\
\href{https://lasair-ztf.lsst.ac.uk/objects/ZTF19aalohjc/}{ZTF19aalohjc}	&	2020afbz	&	SN	&	Non-SN-like light curve	&	2	\\
\href{https://lasair-ztf.lsst.ac.uk/objects/ZTF19aausweg/}{ZTF19aausweg}	&	2019fex	&	SN	&	Non-SN-like light curve	&	2	\\
... & & & \\
\enddata
\tablecomments{Table 3 is published in its entirety in the machine-readable format. A portion is shown here for guidance regarding its form and content.}
\tablenotetext{\dagger}{This object gets flagged by both the object features autoencoder and triplet image autoencoder.}
\tablenotetext{\star}{This object gets flagged by both the object features autoencoder and the light curve autoencoder.}
\end{deluxetable}

\subsection{Triplet image cutouts autoencoder anomalies \label{sec:triplet_anoms}}

Applying the anomaly threshold defined from the test set (described in Section~\ref{sec:algorithm_tc}), the triplet image autoencoder flags 127 objects as anomalous. Using contextual classifications from {\sc sherlock}, these comprise 78 supernovae, 31 AGN, eight orphan transients with no host association, four nuclear transients, four objects with unclear host associations, and two CVs. Alerts flagged with high anomaly scores display reconstruction losses comparable to those measured for the most anomalous objects in the held-out test set.

In contrast to the object features autoencoder, the triplet image autoencoder identifies a relatively homogeneous class of anomalous objects. High anomaly scores are predominantly associated with sources that have faint, ambiguous, or undetected host galaxies. As noted previously, these observational characteristics are commonly associated with rare and luminous transients, including SLSNe; however, this interpretation cannot be confirmed from alert-level image data alone and requires spectroscopic follow-up. 

We perform a manual inspection of the supernova-labeled anomalies to assess their scientific interest. These objects are listed in Table~\ref{tab:tc_anoms1}, and are discussed in detail below.

\subsubsection{Most interesting objects: rank 3}

Objects assigned rank~3 are long-lived transients with faint, unclear, or no hosts. We assign rank~3 to 15 objects, of which only two overlap with the object features anomaly set; the remainder are unique to the image based analysis. This limited overlap indicates that the two approaches are sensitive to complementary manifestations of anomalous behavior. Spectroscopic follow-up is required to determine the physical nature of these objects, which are unlikely to be representative of typical supernovae.

Several rank 3 objects have additional contextual information that supports these interpretations. The coordinates of AT~2024adaw coincide with a known BL Lac object \citep{DAbrusco2019}, consistent with AGN variability, although ALeRCE predicts either a SLSN or a Type II supernova classification. AT~2020qzc coincides with an X-ray source \citep{Voges2000}, supporting an AGN or TDE interpretation. AT~2021vgt, AT~2021usy, and AT~2024quh are classified by ALeRCE as SLSN candidates. AT~2024qs and AT~2022fvo were independently identified by the object features autoencoder and discussed in Section~\ref{sec:feature_anoms}.

\subsubsection{Moderately interesting objects: rank 2}

Objects assigned rank 2 display shorter-lived variability than rank 3 objects, but exhibit light curve morphologies inconsistent with typical supernova evolution. These objects also preferentially lack clear host galaxies or are associated with very faint hosts. Only one rank~2 object is independently identified by the object features autoencoder. Additional contextual information is available for two objects in this category. The coordinates of AT~2024igr coincide with a quasar \citep{Paris2018}, and ALeRCE predicts this object to be an AGN. AT~2025owq is predicted by ALeRCE to be a quasar.

\subsubsection{Least interesting objects: rank 1}

Objects assigned rank 1 are not prioritized for further analysis. These sources are primarily flagged as anomalous due to elevated noise levels in the triplet image cutouts rather than intrinsically unusual astrophysical behavior. Most exhibit either typical light curve evolution or are sparsely sampled, which limits the scope for physical interpretation within the alert stream.

\begin{deluxetable}{lllll}
\tablecaption{Alert stream anomalies flagged by the triplet image autoencoder}
\label{tab:tc_anoms1}
\tabletypesize{\scriptsize}
\tablehead{
\colhead{ZTF ID} & \colhead{IAU Name} & \colhead{ {\sc sherlock} Classification} & \colhead{Notes} & \colhead{Rank}
}
\startdata
\href{https://lasair-ztf.lsst.ac.uk/objects/ZTF20abpwtmi/}{ZTF20abpwtmi} 	&	2020qzc	&	UNCLEAR	&	800 day evolution. Unclear host. Non-SN-like light curve	&	3	\\
\href{https://lasair-ztf.lsst.ac.uk/objects/ZTF24aaivize/}{ZTF24aaivize} 	&	2024ghi	&	SN	&	800 day rise. Faint host. Non-SN-like light curve	&	3	\\
\href{https://lasair-ztf.lsst.ac.uk/objects/ZTF21abrpvho/}{ZTF21abrpvho}	&	2021vgt	&	SN	&	800 day evolution. Faint host. Non-SN-like light curve	&	3	\\
\href{https://lasair-ztf.lsst.ac.uk/objects/ZTF22aagawty/}{ZTF22aagawty\tablenotemark{$\dagger$}}	&	2022fvo	&	SN	&	800 day evolution. Faint host. Non-SN-like light curve	&	3	\\
\href{https://lasair-ztf.lsst.ac.uk/objects/ZTF22aadnwpu/}{ZTF22aadnwpu}	&	2022fvo	&	SN	&	800 day evolution. Faint host. Non-SN-like light curve	&	3	\\
\href{https://lasair-ztf.lsst.ac.uk/objects/ZTF21abqwtvp/}{ZTF21abqwtvp}	&	2021usy	&	SN	&	800 day evolution. Faint host. Non-SN-like light curve	&	3	\\
\href{https://lasair-ztf.lsst.ac.uk/objects/ZTF24aacdofk/}{ZTF24aacdofk}	&	2024elb	&	SN	&	700 day rise. Unclear host. Non-SN-like light curve	&	3	\\
\href{https://lasair-ztf.lsst.ac.uk/objects/ZTF20abnighe/}{ZTF20abnighe}	&	2021vqa	&	SN	&	700 day rise. Non-SN-like light curve	&	3	\\
\href{https://lasair-ztf.lsst.ac.uk/objects/ZTF18aavjoas/}{ZTF18aavjoas}	&	2019hjc	&	SN	&	700 day evolution. Unclear host. Non-SN-like light curve	&	3	\\
\href{https://lasair-ztf.lsst.ac.uk/objects/ZTF23abnkdcu/}{ZTF23abnkdcu}	&	2024pjw	&	SN	&	700 day evolution. Non-SN-like light curve	&	3	\\
\href{https://lasair-ztf.lsst.ac.uk/objects/ZTF21aavhxjc/}{ZTF21aavhxjc}	&	2020iqs	&	SN	&	700 day evolution. Non-SN-like light curve	&	3	\\
\href{https://lasair-ztf.lsst.ac.uk/objects/ZTF23aagdltf/}{ZTF23aagdltf\tablenotemark{$\dagger$}}	&	2024qs	&	SN	&	700 day evolution. Non-SN-like light curve	&	3	\\
\href{https://lasair-ztf.lsst.ac.uk/objects/ZTF23aajibbm/}{ZTF23aajibbm}	&	2025aex	&	SN	&	700 day evolution. Faint host. Non-SN-like light curve	&	3	\\
\href{https://lasair-ztf.lsst.ac.uk/objects/ZTF23abjkwog/}{ZTF23abjkwog}	&	2024quh	&	SN	&	500 day evolution. Faint host. Non-SN-like light curve	&	3	\\
\href{https://lasair-ztf.lsst.ac.uk/objects/ZTF24abuylxw/}{ZTF24abuylxw}	&	2024adaw	&	SN	&	400 day evolution. Non-SN-like light curve	&	3	\\
\href{https://lasair-ztf.lsst.ac.uk/objects/ZTF25aavgvkc/}{ZTF25aavgvkc}	&	2025osj	&	ORPHAN	&	Non-SN-like light curve. No host	&	2	\\
\href{https://lasair-ztf.lsst.ac.uk/objects/ZTF20actkwpz/}{ZTF20actkwpz}	&	2020iov	&	UNCLEAR	&	Non-SN-like light curve.Unclear host	&	2	\\
\href{https://lasair-ztf.lsst.ac.uk/objects/ZTF21aaanzzg/}{ZTF21aaanzzg\tablenotemark{$\dagger$}}	&	2021acf	&	UNCLEAR	&	Non-SN-like light curve. Unclear host	&	2	\\
\href{https://lasair-ztf.lsst.ac.uk/objects/ZTF25aaxfbku/}{ZTF25aaxfbku}	&	2025owq	&	SN	&	Non-SN-like light curve. Faint host	&	2	\\
\href{https://lasair-ztf.lsst.ac.uk/objects/ZTF21aaigcnr/}{ZTF21aaigcnr}	&	2025afas	&	SN	&	Non-SN-like light curve. Faint host	&	2	\\
\href{https://lasair-ztf.lsst.ac.uk/objects/ZTF21aazovbp/}{ZTF21aazovbp}	&	2024jqc	&	SN	&	Non-SN-like light curve. Faint host	&	2	\\
\href{https://lasair-ztf.lsst.ac.uk/objects/ZTF18aaaedbi/}{ZTF18aaaedbi}	&	2025agpv	&	SN	&	Non-SN-like light curve	&	2	\\
\href{https://lasair-ztf.lsst.ac.uk/objects/ZTF18aauhpyb/}{ZTF18aauhpyb\tablenotemark{$\ddagger$}}	&	2018iuh	&	SN	& Non-SN-like light curve	&	2	\\
... & & & \\
\enddata
\tablecomments{Table 4 is published in its entirety in the machine-readable format. A portion is shown here for guidance regarding its form and content.}
\tablenotetext{\dagger}{This object gets flagged by both the object features autoencoder and triplet image autoencoder.}
\tablenotetext{\ddagger}{This object gets flagged by both the triplet image autoencoder and the light curve autoencoder.}
\end{deluxetable}

\subsection{Light curve autoencoder anomalies \label{sec:lightcurve_anoms}}
 
Applying the anomaly threshold defined from the test set (described in Section~\ref{sec:algorithm_light curve}), the light curve autoencoder flags 359 objects as anomalous. Using {\sc sherlock} classifications, these comprise 140 AGN, 139 supernovae, 42 CV, 20 nuclear transients, nine objects with unclear classifications, and nine orphan transients with no associated host. Alerts assigned high anomaly scores exhibit reconstruction losses comparable to those of the most anomalous objects identified in the held-out test set.

The anomalous objects flagged by the light curve autoencoder fall into several broad morphological categories. Some exhibit flat light curves with no well defined peak, while others show very sharp, isolated peaks with little or no sustained evolution on either side. Others display secondary sharp peaks tens to hundreds of days before or after the primary peak.

As in the previous sections, the autoencoder provides a ranked list of anomalous candidates. We then perform a manual inspection of sources classified as supernovae, orphan transients, and objects with unclear classifications, and assign our qualitative priority rank. These objects are listed in Table~\ref{tab:light curve_anoms1} and discussed below.

\subsubsection{Most interesting objects: rank 3}

Rank~3 objects are long-lived transients whose light curve evolution is inconsistent with that of normal supernovae and which lack definitive spectroscopic classification. We assign rank~3 to 60 objects, of which only three overlap with anomalies identified by the object features autoencoder and one with the triplet image autoencoder. A large fraction of rank~3 objects have existing contextual information consistent with CV activity. Several are reported as CV candidates in TNS, such as AT~2017gkd \citep{Delgado2017b}, AT~2020slo \citep{Hodgkin2020}, AT~2021tbw \citep{Hodgkin2021}, AT~2023adwz \citep{Xu2024a}, AT~2018ing, AT~2022aeeu \citep{Hodgkin2022}, and AT~2016dfa. Others have coordinates coincident with known CVs, such as AT~2019qbi \citep{Yecheistov2013}, AT~2019evn \citep{Yecheistov2014}, AT~2019wxe \citep{Gress2015}, AT~2017iwm \citep{Tomasella2016}, AT~2018cvp \citep{Szkody2020}, AT~2021gfx \citep{Kawash2021}, and AT~2024leo \citep{Inight2023}. 

Additional rank~3 objects are associated with other classes of stellar variability. The coordinates of AT~2021twr and AT~2022sfe coincide with classical novae \citep{Samus2017, De2022}. AT~2019cyd is linked to a white dwarf candidate \citep{GentileFusillo2019}. AT~2021ahpz and AT~2017mh are both associated with eclipsing binary classifications \citep{GaiaDR3}. AT~2022ddv is associated with a known young stellar object \citep{Cottar2014}. AT~2021gfx is reported as a variable star on TNS \citep{Tan2021}, but has coordinates coincident with a cataclysmic binary \citep{Kawash2021}. 

A smaller subset of rank~3 objects show indications of extragalactic or nuclear activity. AT~2018gwf has coordinates associated with a radio source \citep{Intema2017} and may correspond to an AGN or TDE. AT~2021pju is predicted as a quasar by ALeRCE, while AT~2021abit is predicted by ALeRCE as a SLSN. AT~2018mkw has been independently flagged as anomalous by the Fink broker. 

AT~2017cox, AT~2021adsf, and AT~2021adir were independently identified by the object features autoencoder and discussed in Section~\ref{sec:feature_anoms}. 

\subsubsection{Moderately interesting objects: rank 2}

Rank~2 objects have not had enough data to let us see their evolution over multi-year time scales, but display abrupt or sharply peaked light curves. Their physical nature is less clear, and continued monitoring may be required to determine whether they represent unusual transients or more common forms of variability.

The algorithm identifies 61 rank~2 objects, of which only three overlap with those flagged by the object features autoencoder, and one overlaps with the triplet image autoencoder.

Several rank~2 objects are associated with CV-like or stellar variability based on contextual information. These include CV candidates reported in TNS, such as  AT~2017cjf \citep{Delgado2017a},  AT~2018ggg \citep{Delgado2018}, AT~2024pic \citep{Xu2024a}, AT~2025adxm \citep{Xu2025b}, and AT~2025abul \citep{Xu2025a}. There are also objects with coordinates coincident with known CVs, such as AT~2021tbt \citep{Gress2014}, AT~2019nuv and AT~2025agcz \citep{Drake2014}, AT~2019wxe \citep{Gress2015}, AT~2020zvs \citep{Szkody2021}, and AT~2019csf \citep{Kawash2021}. AT~2021aavt has coordinates linked to an eclipsing binary \citep{Chen2020}.  AT~2021zie is associated with variability in a known young stellar object \citep{Cutri2003}. 

A subset of rank~2 objects shows indications of extragalactic activity. AT~2022fxt is predicted as AGN by ALeRCE, while the coordinates of AT~2023win coincide with a known quasar \citep{Caccianiga2019}. AT~2025abae is associated with a radio galaxy, suggesting possible AGN-related variability \citep{Ahumada2020}.

AT~2018mkt has been independently flagged as anomalous by the Fink broker, reinforcing its classification as a candidate of interest. AT~2018hgp, AT~2023cqq and AT~2016hfm have been independently identified by the object features autoencoder and discussed in Section~\ref{sec:feature_anoms}.

\subsubsection{Lowest-priority objects: rank 1}

Rank~1 objects display non-standard light curve structure, but have already been spectroscopically classified. This group includes spectroscopically confirmed two SN~II, two SN~Ic, nine SN~Ia, two SLSNe, one Nova and four CVs. Although their light curves are atypical, they are not prioritized for further follow-up within this framework, as their nature is already established.

\begin{deluxetable}{lllll}
\tablecaption{Alert stream anomalies flagged by the light curve autoencoder}
\label{tab:light curve_anoms1}
\tabletypesize{\scriptsize}
\tablehead{
\colhead{ZTF ID} & \colhead{IAU Name} & \colhead{ {\sc sherlock} Classification} & \colhead{Notes} & \colhead{Rank}
}
\startdata
\href{https://lasair-ztf.lsst.ac.uk/objects/ZTF18aaqikyu/}{ZTF18aaqikyu}	&	2018gir	&	SN	&	800 day evolution. Unclear host. Non-SN-like light curve	&	3	\\
\href{https://lasair-ztf.lsst.ac.uk/objects/ZTF18abviaof/}{ZTF18abviaof}	&	2023adwz	&	UNCLEAR	&	800 day evolution. Unclear host. Non-SN-like light curve	&	3	\\
\href{https://lasair-ztf.lsst.ac.uk/objects/ZTF18acmwuum/}{ZTF18acmwuum}	&	2018ing	&	SN	&	800 day evolution. Non-SN-lilke light curve	&	3	\\
\href{https://lasair-ztf.lsst.ac.uk/objects/ZTF18adlqbdc/}{ZTF18adlqbdc}	&	2019evn	&	SN	&	800 day evolution. Non-SN-like light curve	&	3	\\
\href{https://lasair-ztf.lsst.ac.uk/objects/ZTF18aaqzomm/}{ZTF18aaqzomm}	&	2019cyd	&	SN	&	800 day evolution. Non-SN-like light curve	&	3	\\
\href{https://lasair-ztf.lsst.ac.uk/objects/ZTF18acswncz/}{ZTF18acswncz}	&	2022aeeu	&	SN	&	800 day evolution. Non-SN-like light curve	&	3	\\
\href{https://lasair-ztf.lsst.ac.uk/objects/ZTF17aadfiyg/}{ZTF17aadfiyg}	&	2020slo	&	SN	&	800 day evolution. Non-SN-like light curve	&	3	\\
\href{https://lasair-ztf.lsst.ac.uk/objects/ZTF18adkwcxh/}{ZTF18adkwcxh}	&	2021gfx	&	SN	&	800 day evolution. Non-SN-like light curve	&	3	\\
\href{https://lasair-ztf.lsst.ac.uk/objects/ZTF18aboehhn/}{ZTF18aboehhn}	&	2018fjk	&	SN	&	800 day evolution. Non-SN-like light curve	&	3	\\
\href{https://lasair-ztf.lsst.ac.uk/objects/ZTF18admcdum/}{ZTF18admcdum}	&	2019wxe	&	SN	&	800 day evolution. Non-SN-like light curve	&	3	\\
\href{https://lasair-ztf.lsst.ac.uk/objects/ZTF18acjnusv/}{ZTF18acjnusv}	&	2018imh	&	SN	&	800 day evolution. Non-SN-like light curve	&	3	\\
\href{https://lasair-ztf.lsst.ac.uk/objects/ZTF18abgjrmo/}{ZTF18abgjrmo}	&	2021gfx	&	SN	&	800 day evolution. Non-SN-like light curve	&	3	\\
\href{https://lasair-ztf.lsst.ac.uk/objects/ZTF18abxyuzo/}{ZTF18abxyuzo}	&	2018gue	&	SN	&	800 day evolution. Non-SN-like light curve	&	3	\\
\href{https://lasair-ztf.lsst.ac.uk/objects/ZTF18abmpksk/}{ZTF18abmpksk}	&	2016dfa	&	SN	&	800 day evolution. Non-SN-like light curve	&	3	\\
\href{https://lasair-ztf.lsst.ac.uk/objects/ZTF19abdkbtm/}{ZTF19abdkbtm}	&	2018glt	&	SN	&	800 day evolution. Non-SN-like light curve	&	3	\\
\href{https://lasair-ztf.lsst.ac.uk/objects/ZTF18abxcues/}{ZTF18abxcues}	&	2018gwf	&	SN	&	800 day evolution. Non-SN-like light curve	&	3	\\
\href{https://lasair-ztf.lsst.ac.uk/objects/ZTF18aazffjy/}{ZTF18aazffjy}	&	2018cvp	&	SN	&	800 day evolution. Non-SN-like light curve	&	3	\\
\href{https://lasair-ztf.lsst.ac.uk/objects/ZTF18abotssb/}{ZTF18abotssb}	&	2017fuz	&	SN	&	800 day evolution. Non-SN-like light curve	&	3	\\
\href{https://lasair-ztf.lsst.ac.uk/objects/ZTF18adkbcsk/}{ZTF18adkbcsk}	&	2017gkd	&	SN	&	800 day evolution. Non-SN-like light curve	&	3	\\
\href{https://lasair-ztf.lsst.ac.uk/objects/ZTF18acbvqwb/}{ZTF18acbvqwb}	&	2018iwt	&	SN	&	800 day evolution. Non-SN-like light curve	&	3	\\
\href{https://lasair-ztf.lsst.ac.uk/objects/ZTF18abxpvfe/}{ZTF18abxpvfe}	&	2018cuo	&	SN	&	800 day evolution. Non-SN-like light curve	&	3	\\
\href{https://lasair-ztf.lsst.ac.uk/objects/ZTF17aabpjjg/}{ZTF17aabpjjg}	&	2021tbw	&	SN	&	800 day evolution. Non-SN-like light curve	&	3	\\
\href{https://lasair-ztf.lsst.ac.uk/objects/ZTF22abazrjk/}{ZTF22abazrjk}	&	2022sfe	&	SN	&	800 day evolution. Non-SN-like light curve	&	3	\\
\href{https://lasair-ztf.lsst.ac.uk/objects/ZTF21abmbzax/}{ZTF21abmbzax}	&	2021twr	&	SN	&	800 day evolution. Non-SN-like light curve	&	3	\\
\href{https://lasair-ztf.lsst.ac.uk/objects/ZTF18abnyiou/}{ZTF18abnyiou}	&	2024leo	&	SN	&	800 day evolution. Non-SN-like light curve	&	3	\\
\href{https://lasair-ztf.lsst.ac.uk/objects/ZTF18actyzxn/}{ZTF18actyzxn}	&	2022ddv	&	SN	&	800 day evolution. Non-SN-like light curve	&	3	\\
\href{https://lasair-ztf.lsst.ac.uk/objects/ZTF17aachnrk/}{ZTF17aachnrk}	&	2018iqr	&	SN	&	800 day evolution. Non-SN-like light curve	&	3	\\
\href{https://lasair-ztf.lsst.ac.uk/objects/ZTF18abjlyev/}{ZTF18abjlyev}	&	2016eph	&	SN	&	800 day evolution. Non-SN-like light curve	&	3	\\
\href{https://lasair-ztf.lsst.ac.uk/objects/ZTF18adlynmw/}{ZTF18adlynmw}	&	2019hfp	&	SN	&	800 day evolution. Non-SN-like light curve	&	3	\\
\href{https://lasair-ztf.lsst.ac.uk/objects/ZTF21acjcarx/}{ZTF21acjcarx\tablenotemark{$\star$}}	&	2021adsf	&	SN	&	800 day evolution. Non-SN-like light curve	&	3	\\
\href{https://lasair-ztf.lsst.ac.uk/objects/ZTF18abslbyi/}{ZTF18abslbyi}	&	2018fwh	&	SN	&	800 day evolution. Non-SN-like light curve	&	3	\\
\href{https://lasair-ztf.lsst.ac.uk/objects/ZTF20abofezp/}{ZTF20abofezp}	&	2020boi	&	SN	&	800 day evolution. Far from host. Non-SN-like light curve	&	3	\\
\href{https://lasair-ztf.lsst.ac.uk/objects/ZTF18abtsmgm/}{ZTF18abtsmgm}	&	2018fyj	&	SN	&	800 day evolution. Far from host. Non-SN-like light curve	&	3	\\
\href{https://lasair-ztf.lsst.ac.uk/objects/ZTF21abflgvs/}{ZTF21abflgvs}	&	2021pju	&	SN	&	800 day evolution. Far from host. Non-SN-like light curve	&	3	\\
\href{https://lasair-ztf.lsst.ac.uk/objects/ZTF18aaxzbyo/}{ZTF18aaxzbyo}	&	2017mh	&	SN	&	800 day evolution. Far from host. Non-SN-like light curve	&	3	\\
\href{https://lasair-ztf.lsst.ac.uk/objects/ZTF18abjcyha/}{ZTF18abjcyha}	&	2018gwh	&	SN	&	800 day evolution. Far from host. Non-SN-like light curve	&	3	\\
\href{https://lasair-ztf.lsst.ac.uk/objects/ZTF18abtmazz/}{ZTF18abtmazz}	&	2018mkw	&	SN	&	800 day evolution. Far from host. Non-SN-like light curve	&	3	\\
\href{https://lasair-ztf.lsst.ac.uk/objects/ZTF18abcleke/}{ZTF18abcleke}	&	2018gqp	&	SN	&	800 day evolution. Faint host. Non-SN-like light curve	&	3	\\
\href{https://lasair-ztf.lsst.ac.uk/objects/ZTF18adjccrf/}{ZTF18adjccrf}	&	2021njp	&	SN	&	800 day evolution. Faint host. Non-SN-like light curve	&	3	\\
\href{https://lasair-ztf.lsst.ac.uk/objects/ZTF21achjkys/}{ZTF21achjkys}	&	2021abit	&	SN	&	800 day evolution. Faint host. Non-SN-like light curve	&	3	\\
\href{https://lasair-ztf.lsst.ac.uk/objects/ZTF18abjrdjs/}{ZTF18abjrdjs}	&	2018lqm	&	SN	&	800 day evolution. Faint host. Non-SN-like light curve	&	3	\\
\href{https://lasair-ztf.lsst.ac.uk/objects/ZTF18abssoen/}{ZTF18abssoen}	&	2018kcx	&	SN	&	800 day evolution. Faint host. Non-SN-like light curve	&	3	\\
\href{https://lasair-ztf.lsst.ac.uk/objects/ZTF18admwica/}{ZTF18admwica}	&	2021lna	&	SN	&	800 day evolution. Faint host. Non-SN-like light curve	&	3	\\
\href{https://lasair-ztf.lsst.ac.uk/objects/ZTF18admfbxu/}{ZTF18admfbxu}	&	2024aaga	&	SN	&	800 day evolution. Faint host. Non-SN-like light curve	&	3	\\
\href{https://lasair-ztf.lsst.ac.uk/objects/ZTF19aavprpy/}{ZTF19aavprpy}	&	2018cyo	&	SN	&	800 day evolution. Faint host. Non-SN-like light curve	&	3	\\
\href{https://lasair-ztf.lsst.ac.uk/objects/ZTF18abtzkab/}{ZTF18abtzkab}	&	2018gdp	&	SN	&	800 day evolution. Faint host. Non-SN-like light curve	&	3	\\
\href{https://lasair-ztf.lsst.ac.uk/objects/ZTF19aabbgoy/}{ZTF19aabbgoy}	&	2019qbi	&	SN	&	700 day evolution. Non-SN-like light curve	&	3	\\
\href{https://lasair-ztf.lsst.ac.uk/objects/ZTF18aaqdozs/}{ZTF18aaqdozs}	&	2018bms	&	SN	&	700 day evolution. Non-SN-like light curve	&	3	\\
\href{https://lasair-ztf.lsst.ac.uk/objects/ZTF18adovdni/}{ZTF18adovdni}	&	2018bbh	&	SN	&	700 day evolution. Non-SN-like light curve	&	3	\\
... & & & \\
\enddata
\tablecomments{Table 5 is published in its entirety in the machine-readable format. A portion is shown here for guidance regarding its form and content.}
\tablenotetext{\star}{This object gets flagged by both the object features autoencoder and the light curve autoencoder.}
\end{deluxetable}

\section{Discussion and Conclusion \label{sec:conclusion}}

In this work, we apply unsupervised autoencoder based anomaly detection to three complementary alert stream data modalities provided by the Lasair broker: object features, triplet image cutouts, and light curves. We develop a separate autoencoder for each modality and evaluate their performance using both a spectroscopically classified data set and the live ZTF alert stream. 

Treating each modality independently is motivated by practical considerations. In practice, alert packets are frequently incomplete. Some alerts lack derived object features, while others lack image cutouts or sufficiently sampled light curves. A single combined model would therefore exclude a substantial fraction of alerts. In tests of such a combined approach, we found that approximately half of nightly alerts were discarded due to missing modality information, resulting in a significant loss of potentially interesting events. By contrast, the ensemble approach maximizes alert stream coverage and ensures that objects are not excluded solely due to missing data products. 

We evaluate the performance of our algorithms using two metrics: the recall of exotic objects and the purity of anomalous objects, where anomalous objects include both exotic transients and supernova-labeled sources with anomalous properties. Both metrics are evaluated at rank k. For the test set, rank k corresponds to the top ten objects with the highest reconstruction loss. For the alert stream, rank k is defined by the reconstruction loss threshold adopted for each algorithm. The recall and purity metrics for each algorithm are reported in Table~\ref{tab:metrics}.

\begin{deluxetable}{lcc}
\tablecaption{Evaluation metrics of each algorithm in the test set and alert stream at rank $k$.}
\label{tab:metrics}
\tablehead{
\colhead{} & \colhead{Exotic Recall} & \colhead{Purity}
}
\startdata
Object features test set        & 2.3\% & 70\% \\
Object features alert stream    & 11\%  & 65\% \\
Triplet image test set         & 10\%  & 40\% \\
Triplet image alert stream     & 4.9\% & 47\% \\
Light curve test set           & 8.3\% & 60\% \\
Light curve alert stream       & 24\%  & 97\% \\
\enddata
\tablecomments{For the test set, rank $k$ corresponds to the top ten objects with the highest reconstruction loss in the held-out sample. For the alert stream, rank $k$ is defined by the anomaly threshold adopted for each algorithm.}
\end{deluxetable}

In the test set, exotic objects are identified using spectroscopic classifications. In the alert stream, exotic objects are identified using contextual classifications provided by {\sc sherlock}. We note that {\sc sherlock} classifications are not always correct, and therefore the alert stream metrics should be interpreted as indicative rather than definitive measures of performance. They are intended to provide a qualitative assessment of how each algorithm behaves in a realistic alert stream setting, rather than a fully controlled evaluation. We now summarize the performance of each algorithm on the test set and the alert stream.

The object features autoencoder is sensitive to rapid changes in photometry and host association properties. In the test set, it recovers both exotic objects and sources with faint, absent, or incorrect host associations, as well as objects with missing redshift information. For the test set, the purity and exotic object recall are 70\% and 2.3\%, respectively. When applied to the live alert stream, the algorithm flags 207 objects as anomalous. Of these, 103 are labeled by {\sc sherlock} as non-supernova objects and lack spectroscopic classification, corresponding to an exotic object recall of 11\% based on {\sc sherlock} labels. In addition, 32 objects labeled by {\sc sherlock} as supernovae exhibit anomalous properties and are considered suitable for follow-up. Overall, 135 of the 207 flagged objects are considered anomalous, yielding a purity of 65\%. 

The triplet image autoencoder focuses on transient morphology in image cutouts. Overall, the image autoencoder reliably reconstructs typical galaxy structure and low noise difference images, while assigning high anomaly scores to sources with noisy cutouts, faint or unresolved hosts, or ambiguous subtraction residuals. In the held-out test set, the most anomalous objects are those with faint or absent host galaxies. For this test set, the algorithm achieves a purity of 40\% and an exotic object recall of 10\%. When applied to the alert stream, the algorithm flags 127 objects as anomalous. Of these, 37 are labeled by {\sc sherlock} as exotic objects and lack spectroscopic classification, corresponding to an exotic object recall of 4.9\%. Among the remaining 90 objects that are unlabeled or labeled as supernovae, 23 are considered anomalous and recommended for follow-up, yielding a purity of 47\%. The remaining flagged objects mainly have faint, unclear, or absent host galaxies.

The light curve autoencoder plays a complementary role by operating in a restricted regime. The autoencoder is trained only on transients whose evolution occurs on timescales typical of normal supernovae. In the held-out test set, the most extreme anomalies correspond to objects whose light curve evolution is inconsistent with supernova behavior, including sources with multi-year, slowly evolving or non-declining light curves, double-peaked morphologies, or poorly defined maxima. The purity and exotic object recall achieved in the held-out test set for the top ten anomalies are 60\% and 8/3\% respectively. When applied to the live alert stream, the light curve autoencoder flags 361 objects as anomalous. Of these, 220 are not labeled as supernovae by {\sc sherlock}. Among the 141 supernova-labeled objects, 121 are considered sufficiently interesting to warrant follow-up. 20 supernovae labeled objects in this set have already been spectroscopically classified, with seven of them being two SLSNe, one Nova and four CVs and requiring an updated classification by {\sc sherlock}. Therefore the purity of this set is 97\% and the exotic object recall is 24\%.

\begin{figure}
    \centering
    \includegraphics[width=\linewidth]{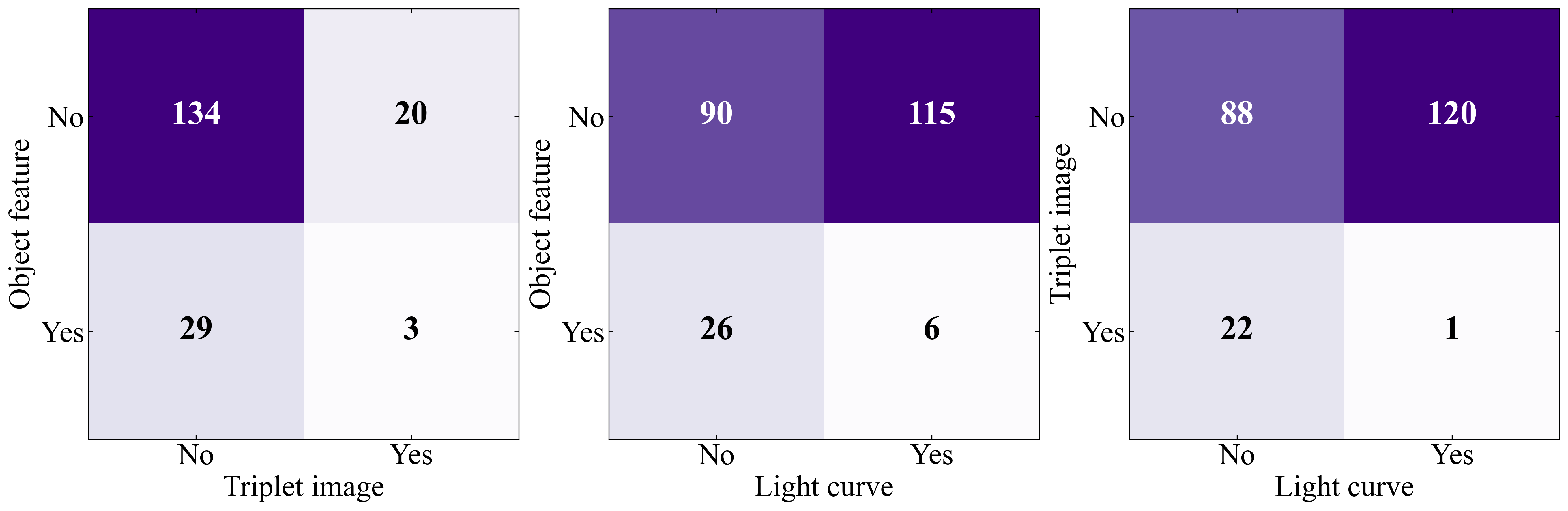}
    \caption{Agreement between anomaly detection algorithms for supernova-labeled objects. Each panel shows a $2\times2$ matrix comparing follow-up decisions for objects classified as supernovae by {\sc sherlock} and manually inspected by us. For each pair of autoencoder algorithms, the axes indicate whether an object was considered interesting enough for follow-up (rank~3 or rank~2) or not interesting by each method. The numbers in each cell give the count of objects in each category. Objects in the lower-right cells are identified as interesting by both algorithms, while off-diagonal cells correspond to objects selected by only one algorithm. There is no overlap in interesting objects between all three autoencoders. The small number of objects in the shared “Yes-Yes” cells demonstrates that the different data modalities highlight distinct sets of interesting supernova candidates.}
    \label{fig:overlap}
\end{figure}

Across all three autoencoders and data modalities, in the ZTF alert stream spanning 1 to 25 December 2025, we recovered 313 objects classified as supernovae by {\sc sherlock} that required manual inspection by us. Of these, 175 objects were selected and recommended for follow-up. Between the object features and triplet image autoencoders, eleven objects overlap: three AGN, one unclear classification object, and seven supernovae, of which only three are recommended for follow-up. Between the object features and light curve autoencoders, sixteen objects overlap, comprising three CVs, four AGN, and nine supernovae, with six supernovae recommended for follow-up. Between the triplet image and light curve autoencoders, six objects overlap: one CV, three AGN, one orphan transient, and one supernova, the latter being a rank~3 object. No supernova-labeled object is recovered by all three autoencoders; the only object common to all three autoencoders is classified by {\sc sherlock} as an AGN. Overlap between the supernova-labeled anomaly sets recovered by different modalities is illustrated in Figure~\ref{fig:overlap}. These results demonstrate that the algorithms largely recover distinct sets of anomalous objects.

Overall, within 25 days of applying our pipeline to the live alert stream, we obtain 87 rank~3 candidates. We recommend that these objects be prioritized for spectroscopic follow-up, as they may correspond to known exotic populations or previously unrecognized classes of transients. Rank~3 objects across all modalities are characterized by long-lived or persistent signals, which can plausibly arise from several distinct physical scenarios. For example, supernovae may exhibit broad P-Cygni profiles during the photospheric phase or nebular emission lines at late times \citep{Jerkstrand2017}, although the persistence of such features over multi-year timescales would be unexpected. Alternatively, some candidates may be consistent with PISNe, which result from the explosive disruption of very massive stars following electron-positron pair production in the stellar core \citep{Fryer2001}. A further possibility is ambiguous nuclear transients, including well-studied examples such as “Scary Barbie” \citep{Subrayan2023}. Discriminating between these scenarios requires spectroscopic observations spanning ultraviolet to near-infrared wavelengths. Follow-up of these sources would also yield clean, well-characterized samples for future anomaly detection studies.

Although this study uses ZTF data, the Lasair data products closely mirror those expected from the Vera C.\ Rubin Observatory alert stream. Rubin will deliver deeper observations, additional photometric bands, and a substantially higher alert rate, thereby increasing sensitivity to previously unobserved transient phenomena. Transferring this framework to Rubin will require the construction of a training set from early alert data, ideally anchored by spectroscopic classifications, following the same methodology adopted here. We anticipate a higher yield of rank~3 candidates from Rubin, contingent on the availability of a clean and native training set. One potential challenge during early Rubin operations is the availability of well-established reference images over large areas of the sky, which is likely to introduce additional artifacts and spurious detections. This may represent a limiting factor beyond the sheer volume of alerts.

In summary, we show that unsupervised autoencoders are effective for anomaly detection in transient alert streams, and that a curated training set of a few thousand objects is sufficient to recover genuinely unusual objects in both test data and live operation. Treating alert stream modalities independently, rather than combining them into a single model, substantially improves coverage and yields distinct anomalies. Moreover, the ensemble approach enables the identification of both extreme outliers and subtle, high-dimensional deviations that are not evident in low-dimensional projections, providing a broader discovery space. We therefore conclude that an ensemble of modality-specific anomaly detectors, as implemented in our Anomaly Hunter for Alerts ({\sc AHA}) pipeline, offers a scalable and transferable solution for current and next-generation time-domain surveys, where rapid identification of rare and unexpected phenomena is increasingly critical.

\section*{Acknowledgments}
L.I. and S.C. are supported by funding from Breakthrough Listen. The Breakthrough Prize Foundation funds the Breakthrough Initiatives, which manages Breakthrough Listen. C.J.L. acknowledges funding from STFC via LSST:UK. H.F.S. is supported by Schmidt Sciences through the Schmidt AI in Science Fellowship. Thanks to Stephen Smartt and Roy Williams for advice on Lasair transient data retrieval. Lasair is supported by the UKRI Science and Technology Facilities Council and is a collaboration between the University of Edinburgh (grant ST/N002512/1) and Queen’s University Belfast (grant ST/N002520/1) within the LSST:UK Science Consortium.

\bibliographystyle{aasjournalv7}
\bibliography{ref}

\end{document}